\documentclass[aps,letterpaper,twocolumn,pra,longbibliography,footinbib,floatfix,showpacs,superscriptaddress]{revtex4-1}
\usepackage{amsmath}
\usepackage{amsfonts}
\usepackage{amssymb}
\usepackage[scanall]{psfrag}
\usepackage{psfrag}
\usepackage{graphicx}
\usepackage{color}

\begin{document}
\newcommand{\of}[1]{\left( #1 \right)}
\newcommand{\sqof}[1]{\left[ #1 \right]}
\newcommand{\abs}[1]{\left| #1 \right|}
\newcommand{\avg}[1]{\left< #1 \right>}
\newcommand{\cuof}[1]{\left \{ #1 \right \} }
\newcommand{\bra}[1]{\left < #1 \right | }
\newcommand{\ket}[1]{\left | #1 \right > }
\newcommand{\pil}{\frac{\pi}{L}}
\newcommand{\bx}{\mathbf{x}}
\newcommand{\by}{\mathbf{y}}
\newcommand{\bk}{\mathbf{k}}
\newcommand{\bp}{\mathbf{p}}
\newcommand{\bl}{\mathbf{l}}
\newcommand{\bq}{\mathbf{q}}
\newcommand{\bs}{\mathbf{s}}
\newcommand{\psibar}{\overline{\psi}}
\newcommand{\svec}{\overrightarrow{\sigma}}
\newcommand{\dvec}{\overrightarrow{\partial}}
\newcommand{\bA}{\mathbf{A}}
\newcommand{\bdelta}{\mathbf{\delta}}
\newcommand{\bK}{\mathbf{K}}
\newcommand{\bQ}{\mathbf{Q}}
\newcommand{\bG}{\mathbf{G}}
\newcommand{\bw}{\mathbf{w}}
\newcommand{\bL}{\mathbf{L}}
\newcommand{\ohat}{\widehat{O}}
\newcommand{\up}{\uparrow}
\newcommand{\down}{\downarrow}
\newcommand{\MM}{\mathcal{M}}
\author{Eliot Kapit}

\affiliation{Rudolf Peierls Center for Theoretical Physics, University of Oxford, 1 Keble Rd, Oxford, OX1 3NP}

\author{John T. Chalker}

\affiliation{Rudolf Peierls Center for Theoretical Physics, University of Oxford, 1 Keble Rd, Oxford, OX1 3NP}

\author{Steven H. Simon}

\affiliation{Rudolf Peierls Center for Theoretical Physics, University of Oxford, 1 Keble Rd, Oxford, OX1 3NP}

\title{Passive correction of quantum logical errors in a driven, dissipative system: a blueprint for an analog quantum code fabric}

\begin{abstract}

A physical realization of self correcting quantum code would be profoundly useful for constructing a quantum computer. In this theoretical work, we provide a partial solution to major challenges preventing self correcting quantum code from being engineered in realistic devices. We consider a variant of Kitaev's toric code coupled to propagating bosons, which induce a long-ranged interaction between anyonic defects. By coupling the primary quantum system to an engineered dissipation source through resonant energy transfer, we demonstrate a ``rate barrier" which leads to a potentially enormous increase in the system's quantum state lifetime through purely passive quantum error correction, even when coupled to an infinite temperature bath. While our mechanism is not scalable to infinitely large systems,  the maximum effective size can be very large, and it is fully compatible with active error correction schemes. Our model uses only on-site and nearest-neighbor interactions, and could be implemented in superconducting qubits. We sketch one such implementation at the end of this work.

\end{abstract}

\maketitle

\section{Introduction}\label{intro}

The most important obstacles to the construction of a functional quantum computer are noise and decoherence, inescapable disturbances in the quantum state of a system which result from interactions with its environment. A compelling potential solution to this problem is to use systems with topological quantum order (TQO) to encode quantum information nonlocally, which reduces susceptibility to local noise exponentially in increasing system size. In these systems, local noise (e.g. random unitary operations on qubits) cannot distinguish states within a topologically degenerate manifold, and instead, can create fractionalized excitations called anyons. For a noise source to change or measure a quantum state, the anyons must travel a macroscopic, topologically nontrivial path; in the case of systems with closed boundary conditions, this can be visualized as a chain of operations connecting two well-separated edges, and in the case of systems with periodic boundary conditions, the chain of operations must form a non-contractible loop. This requires an extensive number of operations, but as there are unfortunately an extensive number of noise sources, once anyons are created it is surprisingly difficult to ensure that they cannot separate and alter the quantum state.

Approaches to stabilize topological quantum states, then, fall into two broad and non-exclusive categories: mechanisms for pulling anyons back together after they are created, and mechanisms for suppressing the creation of anyons in the first place. Topological error correction codes \cite{kitaev2003,bombin2010,fowlersurface,yaowang} generally fall into the first class: a large number of qubits are arranged in a ``code fabric," where a topological quantum Hamiltonian is simulated digitally through repeated measurements. When anyons are detected by these measurements, a classical algorithm is used to apply a sequence of operations to remove them. Such constructions can be extremely effective when the error rate lies below a given, model-dependent threshold, but the resource costs--- reading out thousands of qubits, evaluating a probabilistic error model, and performing thousands of operations, with the cycle repeated rapidly--- present a daunting experimental challenge.

An alternative avenue of research is the quest for passively stable quantum memory, usually defined to be a topologically ordered system where the energy barrier between ground states (the amount of energy which must be added and removed to toggle between ground states)  grows extensively with system size \cite{bacon2006,hammacastelnovo2009,chesirothlisberger2010,hutterwootton2012,pedrocchihutter2013,bombinchhajlany2013,wootton2013}. Stable \textit{classical} memory is easy to find, with the simplest example being a two dimensional Ising ferromagnet: to flip all the spins from one value to the other, one must create a domain wall that spans the system, which involves an extensive energy cost. As this energy is inaccessible through the coupling to a low temperature bath, the macroscopic magnetization is stable for exponentially long times. However, the 2d Ising model is not a stable quantum memory, since a single local operation can distinguish the two ground states, dephasing a superposition. Unfortunately, the combination of topological order and an extensive energy barrier has proved extremely elusive and no-go theorems \cite{bravyiterhal2009,yoshida2011} have shown that under fairly general sets of assumptions, these features cannot coexist three or fewer spatial dimensions. Induced long range interactions are a potential way around this constraint, and in recent years a number of models have been formulated which introduce long range interactions to exactly solvable ``commuting stabilizer" Hamiltonians with TQO \cite{hammacastelnovo2009,chesirothlisberger2010,hutterwootton2012,pedrocchihutter2013,wootton2013}.

However, we would argue that even models with long ranged interactions are not realistic self correcting quantum memories without further modification. This is because the most feasible implementations of commuting stabilizer models are quantum simulators, where tuned interactions and drive fields are applied to collections of qubits to physically simulate the desired Hamiltonian in a rotating frame. Photonic systems such as superconducting qubits are among the most promising platforms for generating topological states, as the large nonlinearities, tunability and lack of particle number conservation allow for the direct physical construction of exotic Hamiltonians. Because the excitation energies of the individual qubit devices are typically larger than the qubit-qubit interaction energies by more than an order of magnitude, continuous drive fields must be applied to obtain a true many-body system; an example configuration of drive fields which generates topological order in superconducting device arrays is discussed at the end of this work. 

The Hamiltonian of a continuously driven quantum system is almost always evaluated in a rotating frame, but in this case the system's coupling to the outside world is not a simple finite temperature bath. In fact, when considering excitations created by photon losses into the environment, the most realistic error model is usually an infinitely weak coupling to an \textit{infinite} temperature bath, such that the total local flip rate is finite. This occurs because even in the ideal circumstance where the system is surrounded by a zero temperature enclosure, photons can still be spontaneously absorbed by the environment, and as these processes involve physical energy transfers on the order of the qubit energy, small changes in this energy due to many-body effects should not significantly change the rate of escape. Consequently, the height of the energy barrier in the rotating frame Hamiltonian is irrelevant, and the system is not protected against errors induced by its coupling to its environment without further intervention.  

In a recent work \cite{kapithafezi2014}, Kapit, Hafezi and Simon demonstrated that engineered dissipation \cite{diehlmicheli,krausbuchler,verstraete,pastawskikay2009,pastawski,vollbrechtmuschik2011,kastoryanowolf2013,shankarhatridge2013,aronkulkarni2014,mirrahimileghtas2014} can passively stabilize two-dimensional anyonic states of light against particle losses. By introducing an auxiliary ``shadow" lattice of qubits with fast decay rates, hole states created by particle losses can be rapidly refilled in a quantum simulator of the bosonic fractional quantum Hall effect \cite{kapitmueller,kapitgauge,kapitsimon,hafeziadhikari}. This mechanism is very effective for holding the system in or near \textit{a} topological ground state, but does not necessarily protect specific states within the degenerate ground state manifold. Thus, while useful for studying bosonic fractional quantum Hall states, a more complex construction is needed to engineer a useful quantum memory.

In this paper, we further extend the shadow lattice concept to demonstrate a mechanism which induces passive \textit{logical} quantum error correction in driven anyon systems. We begin with a variant of the toric code \cite{kitaev2003} coupled to propagating bosons, which induce a long ranged interaction between commuting stabilizers while still preserving the topological character and exact solution of the original model. We then show that, by adding an engineered dissipation source which is tuned to remove energy very efficiently in specified ranges (determined by the strength of the interaction), we can induce a ``rate barrier" which causes spatially separated pairs of anyons to fall back together much more quickly than the rate of further separation. This leads to an exponential reduction in the quantum logical error rate in increasing system size, and though our mechanism is not scalable to infinitely large systems, the maximum effective size can be large (more than $20 \times 20$ plaquettes in our simulations). Further, error correction in this case is completely passive, and functions independently of any input from an outside observer. Through a series of numerical simulations, we compute the logical error rate as a function of numerous device parameters, observe general trends and provide a loose set of guidelines for optimizing the quantum state lifetime. We also sketch an implementation of this model through circuit QED, and speculate on extending our error correction mechanism to more complex systems with TQO.

The remainder of this paper is organized as follows. In Section~\ref{slec}, we review the shadow lattice mechanism for passive quantum error correction. In Section~\ref{hamsec} we define our Hamiltonian, and reduce it to a purely classical model through a unitary transformation and a few well-controlled approximations. We then introduce a shadow lattice which will pull anyons back together, and provide general guidelines for its construction. Following that, in Section~\ref{ressec}, we present our numerical simulations, and show that the logical error rate can be suppressed exponentially in increasing system size, but beyond a critical size the error correction saturates and ceases to provide further improvements if the system is made any larger. We estimate the length scale at which this occurs and show that it can be large. We present results both for systems with periodic boundary conditions and for systems with edges. We then offer a concluding discussion of our results and future extensions to the model. In the appendices, we summarize the defect tracking algorithm used for our simulations, and propose an example three-body gadget \cite{kempekitaev,jordanfarhi2008,ockoyoshida} to generate the model in superconducting qubits.

\begin{figure}
\includegraphics[width=3.25in]{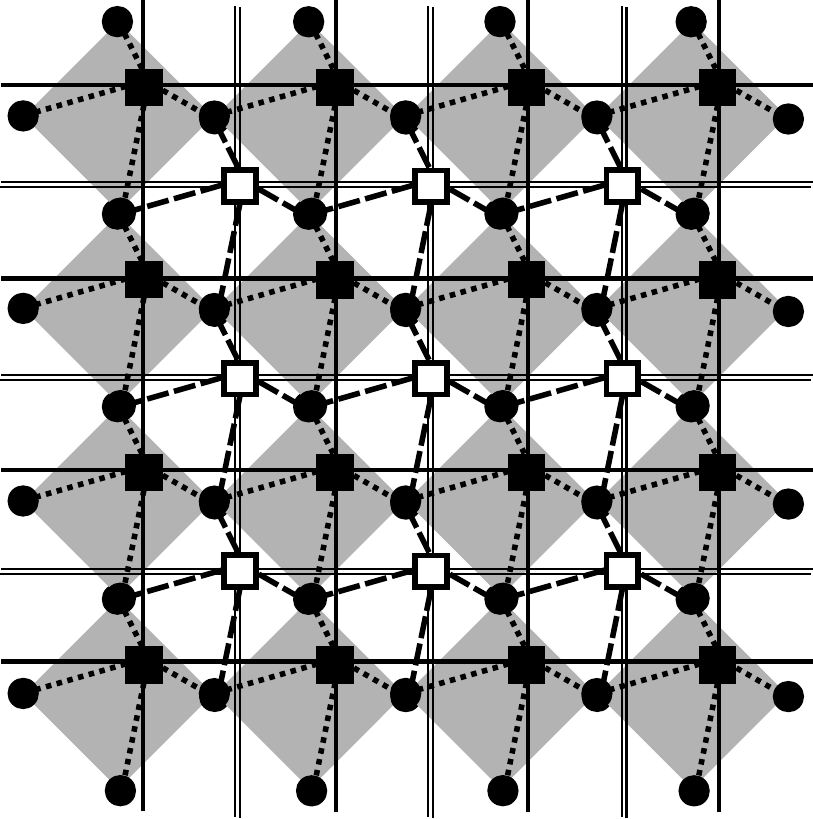}
\caption{Topological qubit array studied in this work, as described in Eq.~\ref{fullH}. Here, the black circles are spins, with the circuitry in the gray shaded regions depicting the 5-body plaquette couplings $W_{\diamondsuit j} \of{V_{A0} + g \of{a_{j}^\dagger + a_j } }$ between the spins and the type $A$ resonators (black filled squares), and the circuitry depicted in the white regions depicting the star couplings $W_{+j} \of{V_{B0} + g \of{b_{j}^\dagger + b_j } }$ to the type $B$ resonators (white squares with black borders). The black single and double lines depict the nearest neighbor hopping between resonators of the same type. The propagating bosons in the resonator arrays induce ranged interactions in the plaquette and star terms. The model can still be solved exactly, and has the same degeneracies and excitation structure of Kitaev's toric code (though there is now an interaction potential between anyons). By coupling the spins to an auxiliary ``shadow" lattice of qubits engineered to have rapid decay rates (not shown in the figure), the system can passively correct quantum logic errors and function as a topologically protected quantum memory. Note that in a realistic device the many-body couplings would have to be engineered through a gadget construction, which requires additional qubits not shown here.}\label{latfig}
\end{figure}

\section{Shadow Lattices and Passive Error Correction}\label{slec}

Before defining the Hamiltonian we will study in this work, we would like to outline our basic mechanism for passive error correction. Specifically, we consider a many-body quantum system which is weakly coupled to an array of intentionally bad qubits or resonators (the ``shadow lattice"), with fast decay rates $\Gamma_S$ and tuned excitation energies $\omega_S$. It is important to note that all quantities are expected to be evaluated in a rotating frame for the underlying physical device, so that the shadow lattice energies $\omega_S$ will often define a detuning from the rotating frame frequency. We demonstrate that unwanted \textit{many-body} excitations can be resonantly transfered to \textit{single particle} states on the shadow lattice, where they rapidly decay, relaxing the combined system back toward its rotating frame ground state without external intervention. As a warmup to the much more complex many-body formulation which forms the subject of this paper, we will review the passive formulation of the 3-qubit bit flip code \cite{reeddicarlo2}, originally proposed in \cite{kapithafezi2014}. 

We consider a system of three qubits, arranged in a ring, which we call the primary lattice. The three qubits are coupled to each other through simple ferromagnetic Ising $xx$ interactions, and each qubit is coupled to a single shadow lattice qubit through a $yy$ ($\sigma_{iP}^{y} \sigma_{iS}^{y}$) interaction. To maintain notational consistency with \cite{kapithafezi2014} we reserve the labels $P$ and $S$ for qubits on the primary and shadow lattices, respectively. Our total system Hamiltonian is:
\begin{eqnarray}\label{Hring}
H &=& -J \of{ \sigma_{1P}^x \sigma_{2P}^x + \sigma_{1P}^x \sigma_{3P}^x + \sigma_{2P}^x \sigma_{3P}^x} \\
& & + \sum_{i=1}^{3} \sqof{ \Omega  \sigma_{iP}^{y} \sigma_{iS}^{y}    + \frac{\omega_S}{2} \sigma_{iS}^{z}  }. \nonumber
\end{eqnarray}
We assume that $J \sim \omega_S \gg \Omega$, and in this limit, the system has a nearly twofold degenerate ground state, where all shadow lattice qubits are in their ground states in the $\sigma^z$ basis, and the three primary qubits are either all in the $\sigma^x = -1$ or all in the $\sigma^x = +1$ state. We call these states $\ket{0_L}$ and $\ket{1_L}$. We now assume that the primary qubits have a random flip rate ($\sigma^y$ errors) of $\Gamma_P$ and the three shadow lattice qubits have a relaxation rate ($\sigma^-$ operations) $\Gamma_S$, where $\Gamma_S \gg \Gamma_P$. Due to this rapid decay, the shadow lattice qubits will nearly always be in their ground states, creating an effective zero temperature bath for the primary system.

Let us assume we begin in state $\ket{0_L}$, and one of the primary lattice bits randomly flips. This creates an excitation of energy $4J$, but due to the coupling to the shadow lattice, a single spin flip is not an eigenstate of the full system Hamiltonian, and instead will flop back and forth between the two lattices. Whenever the excitation occupies the shadow lattice, it can decay rapidly (rate $\Gamma_S$), and if $\omega_S$ is nearly equal to $4J$, then the resonance condition implies that to excite a shadow lattice qubit, an excitation must be removed from the primary lattice. This gives single spin flips on the primary lattice an effective decay rate $\Gamma_R$, which is constrained by the slower of the transfer rate ($\Omega$) to pass the excitation to the shadow lattice and the shadow lattice relaxation rate $\Gamma_S$.

As we are interested in the dynamics of the primary lattice and not the shadow lattice itself, we can integrate out the shadow lattice using Fermi's golden rule to generate the repair rate $\Gamma_R$ for eliminating spin flips on the primary lattice. Repairing a spin flip on the primary lattice is a two-step process, the first (resonant transfer of the excitation to a shadow qubit, at an average rate $\Gamma_{PS}$) of which is reversible, but the second (relaxation of the shadow qubit to its ground state at a rate $\Gamma_S$) is not. The total rate $\Gamma_R \of{\delta E}$ of a process which changes the energy of the primary lattice by $\delta E$ depends on $\Omega$, $\Gamma_S$, and the detuning $\omega_S - 4J$, and is given by:
\begin{eqnarray}\label{Fgolden}
\Gamma_R \of{\delta E} &=& \frac{\Gamma_{PS} \of{\delta E} \times \Gamma_{S}   }{ \Gamma_{PS} \of{\delta E} + \Gamma_S }; \\
\Gamma_{PS} \of{\delta E} &=& 2 \pi \Omega^{2} \rho_S \of{\delta E},  \nonumber \\
\rho_S \of{\delta E} & = & \frac{1}{2 \pi} \frac{ \Gamma_S  }{\of{\delta E - \omega_S}^{2} + \Gamma_{S}^{2} /4 }. \nonumber
\end{eqnarray}
Note also that this calculation predicts the additional error rate induced in the primary lattice by the shadow lattice itself; for $\omega_S = 4J$ the induced error rate is suppressed by a factor of $\Omega^2 / 64 J^2$ compared to the repair rate. The resonance condition thus minimizes errors induced by the shadow lattice while still allowing for rapid bit flip correction.

Since the spin flip on the primary lattice is removed through a second $\sigma^y$ operation, the combined process of a bit flip error and its subsequent correction does not measure the state of the primary lattice, and therefore cannot distinguish the two ground states. If however two spins on the primary lattice flip in quick succession, then it is energetically favorable for the shadow lattice to simply flip the third bit, transferring the state to the other ground state. If we choose $\Gamma_R \gg \Gamma_P$, this process will be rare, and the device's lifetime against bit flips will scale as $\Gamma_R / \Gamma_{P}^{2}$, which can be substantially longer than the lifetime of any single bit on the primary lattice! Thus, by coupling a many-body Hamiltonian to an array of lossy disconnected sites and carefully tuning the energies of the two subsystems, unwanted excitations created by errors in the many-body system can be rapidly eliminated, holding the system in its ground state for long times.

Of course, if a $\sigma^z$ error occurs on the primary lattice, it will be corrected with a $\sigma^y$ operation by the shadow lattice, leading to a combined $\sigma^x$ operation that returns opposite signs for the two ground states. Thus, the relative phase of an arbitrary superposition $\alpha \ket{0_L} + \beta \ket{1_L}$ is not protected against $\sigma^z$ errors (or dephasing from $\sigma^x$ errors), so this device is of limited usefulness for quantum computing purposes. To construct a useful quantum memory based on these principles, the many-body system must display topological order, and the degenerate (logical) ground states must not be distinguishable through local operations. Simultaneously, the shadow lattice must be able to correct extended series of quantum errors that create spatially separated anyons in the primary system; in other words, it must be able to both remove local pairs of topological defects and shrink extended ones, until they once again become local and can be completely eliminated. We will now propose a model which accomplishes both these goals.

\section{The Hamiltonian}\label{hamsec}

\subsection{Definitions}

We consider a square lattice of superconducting qubits (treated here as two level systems) coupled to two interpenetrating lattices of simple resonators $A$ and $B$, as shown in FIG.~\ref{latfig}. The Hamiltonian consists of simple kinetic terms for the propagating bosons, which are coupled to the superconducting qubits through driven plaquette ($\diamondsuit$) and star ($+$) couplings. The lattice contains $N$ plaquettes and $N$ stars. These couplings are products of four $\sigma^x$ ($\sigma^y$) operators acting on all the spins $i$ in a plaquette (star) $j$:
\begin{eqnarray}\label{Wdefs}
W_{\diamondsuit j} & \equiv & \prod_{i \of{ \in \diamondsuit j} = 1}^{4} \sigma_{i}^{x}, \quad W_{+ j} \equiv \prod_{i \of{\in + j} = 1}^{4} \sigma_{i}^{y}
\end{eqnarray}
Note that all the $W_{\diamondsuit}$ and $W_{+}$ operators commute with each other. Each primary qubit is coupled to a collection of shadow lattice qubits (not shown in the figure), which will be described below. Each plaquette couples to a single $A$ site, each star couples to a single $B$ site and each spin is contained in two plaquettes and two stars. We will treat the entire calculation in the rotating frame, so that the Hamiltonian is independent of time; see the appendix for a discussion of this construction. The propagating bosons are noninteracting and do not mix with each other. Our total primary lattice Hamiltonian is
\begin{eqnarray}\label{fullH}
H & = & -J \sum_{\avg{ij} \in A} \of{a_{i}^{\dagger} a_{j} + a_{j}^{\dagger} a_{i}} -J \sum_{\avg{ij} \in B} \of{b_{i}^{\dagger} b_{j} + b_{j}^{\dagger} b_{i}} \nonumber  \\
& & +  \of{4J - \mu_A} \sum_{j \in A} a_{j}^{\dagger} a_j +  \of{4J  - \mu_B} \sum_{j \in B} b_{j}^{\dagger} b_j  \\
& & - \sum_{j \in \diamondsuit} W_{\diamondsuit j} \of{V_{A0} + g \of{a_{j}^\dagger + a_j } } \nonumber \\
& & - \sum_{j \in +} W_{+j} \of{V_{B0} + g \of{b_{j}^\dagger + b_j } }. \nonumber
\end{eqnarray}
If we kept only the plaquette and star terms and ignored the propagating bosons, we would simply have Kitaev's toric code Hamiltonian. Models of this type have been considered previously by multiple authors \cite{hammacastelnovo2009,hutterwootton2012,pedrocchihutter2013,wootton2013}, typically with the goal of demonstrating an extensive energy gap for anyon creation, a property which will not be exhibited by our system, except perhaps in the limit of $\mu_{A/B} \to 0$. $H$ can be diagonalized exactly through a similarity transformation \cite{mahan}. We define $\omega_{A/B \bk} = 4 J - \mu_{A/B} - 2 J \cos \bk_x - 2 J \cos \bk_y$ and:
\begin{eqnarray}
s_{\bk} &\equiv& \sum_{j \in A, \diamondsuit} W_{\diamondsuit j}  \frac{g}{\omega_{A \bk  } \sqrt{N}}  \of{ a_{\bk}^{\dagger} e^{-i \bk \cdot \mathbf{r}_{j} } - a_{\bk} e^{i \bk \cdot \mathbf{r}_{j} }  }  \\
& & + \sum_{j \in B, +} W_{+j}  \frac{g}{\omega_{B \bk }\sqrt{N}}  \of{ b_{\bk}^{\dagger} e^{-i \bk \cdot \mathbf{r}_{j} } - b_{\bk} e^{i \bk \cdot \mathbf{r}_{j} }  }; \nonumber \\
s & \equiv & \sum_{\bk} s_{\bk}. \nonumber
\end{eqnarray}
We note that $s^{\dagger} = -s$. Recalling that all the $W$ operators commute with each other, and invoking the Baker-Campbell-Hausdorff formula,
\begin{eqnarray}
e^{X} Y e^{-X} &=& Y + \sum_{n=1}^{\infty} \sqof{X,Y}^{(n)}, \\
 \sqof{X,Y}^{(1)} &\equiv & \sqof{X,Y}, \quad \sqof{X,Y}^{(n)} \equiv \sqof{X, \sqof{X,Y}^{(n-1)} }, \nonumber
\end{eqnarray}
we see that the transformed Hamiltonian $\tilde{H} \equiv e^{s} H e^{-s}$, up to a constant shift, can be written exactly as:
\begin{eqnarray}\label{tildeH}
\tilde{H} &=& \sum_{\bk} \omega_{A\bk} a_{\bk}^{\dagger} a_{\bk} + \sum_{\bk} \omega_{B\bk} b_{\bk}^{\dagger} b_{\bk} \\
& & - V_{A0} \sum_{j \in \diamondsuit} W_{\diamondsuit j} - \sum_{ij \in \diamondsuit} U_A \of{\mathbf{r}_i - \mathbf{r}_j } W_{\diamondsuit j} W_{\diamondsuit i} \nonumber \\
& & - V_{B0} \sum_{j \in +} W_{+j} - \sum_{ij \in +} U_B \of{\mathbf{r}_i - \mathbf{r}_j } W_{+j} W_{+i}, \nonumber \\
& & U_{A/B} \of{\mathbf{r}_i - \mathbf{r}_j } = \frac{g^{2}}{N} \sum_{\bk} \frac{ e^{i \bk \cdot \of{\mathbf{r}_{i} - \mathbf{r}_{j} }} }{ \omega_{ A/B \bk} }. 
\end{eqnarray}
With this unitary transformation, the spins and propagating bosons are decoupled, and the model can be exactly solved, since the interaction potential $U_{A/B} \of{r}$ does not interfere with the commutation relations that lead to the toric code's exact solution. For $\mu<0$ the potential $U_{A/B} \of{r}$ is positive definite, increases as $\log 1/r$ when $r$ is small and decays as $e^{-\sqrt{-\mu/J} r}$ when $r$ is large. We assume $\mu_A, \mu_B <0$ to avoid any possible instabilities. We are now, finally, in a position where the system can be treated completely classically. Defining $Q_{\diamondsuit/+ j} \equiv \of{1-W_{\diamondsuit/+ j}}/2$, we can rewrite the system as (again dropping a constant zero point energy):
\begin{eqnarray}\label{HC}
\tilde{H} & = & \sum_{\bk} \omega_{A\bk} a_{\bk}^{\dagger} a_{\bk} + \sum_{\bk} \omega_{B\bk} b_{\bk}^{\dagger} b_{\bk} + H_c, \\
H_c &=&   \of{ V_A \sum_{j \in \diamondsuit} Q_{\diamondsuit j} + V_B \sum_{j \in +} Q_{+ j} } \nonumber \\
& & - 8 \sum_{i < j \in \diamondsuit} Q_{\diamondsuit j} Q_{\diamondsuit j} U_{A} \of{\mathbf{r}_i - \mathbf{r}_j } \nonumber \\
& & - 8 \sum_{i < j \in +} Q_{+j} Q_{+j} U_{B} \of{\mathbf{r}_i - \mathbf{r}_j }; \nonumber \\
V_{A/B} & \equiv & 2 V_{A/B0} + 4 \sum_{\mathbf{r \neq 0}} U_{A/B} \of{\mathbf{r}}. \nonumber
\end{eqnarray}
Since $U$ is positive definite, the ground state of $H_{c}$ is simply one where all $W$ operators evaluate to +1 (all $Q$ operators evaluating to zero), and has a degeneracy controlled by the system's boundary conditions. As all the $W$ operators commute with each other, and plaquette or star violations do not propagate (see below), $H_c$ can be simulated as a purely classical model, a fact which will be vital to our quantum error simulations later on. 

\subsection{Excitations and incoherent dynamics}

The ground state of $\tilde{H}$ can be described trivially: since all the $W_\diamondsuit$ and $W_+$ operators commute, and the interaction term is purely attractive, the exact ground state(s) are simply configurations where all $W_\diamondsuit$ and $W_+$ operators evaluate to 1, and where there are no $a$ or $b$ bosons present. Depending on the choice of boundary conditions, this ground state may be degenerate, and we will use this degeneracy to encode quantum information. The excitations of this system are identified by flipped plaquette or star operators ($W_{\diamondsuit/+} = -1$), which are created in pairs, as each spin flip affects two adjoining plaquettes or stars. These defects are anyons, in that winding a plaquette violation in a closed loop around a star violation yields a sign flip of -1 acting on the quantum state. Notably, they do not propagate on their own; as all the terms in the Hamiltonian (\ref{HC}) commute with each other, an anyon can only move from a plaquette/star to one of its neighbors through a spin flip, induced by a coupling to an external system. Logical errors occur in systems with periodic boundary conditions when an anyon pair is created, separates, winds around the system and then recombines, forming a non-contractible loop upon annihilation. The case of edges will be discussed later in this work.

Let us now consider an error, represented by a random $\sigma_{i}^{y}$ operation. The primary source of such errors is from photon losses into the environment, which involve an energy transfer of $\sim \omega_Q$ (the bare excitation energy of the qubits; see appendix B) in the \textit{rest} frame, and thus occur at rates roughly independent of the energetics of the many-body system. Recalling that we transformed all operators $O$ by $O \to e^{s} O e^{-s}$ to diagonalize the Hamiltonian, the $y$ error becomes:
\begin{eqnarray}\label{acty}
e^{s} \sigma_{i}^{y} e^{-s} & = & \sigma_{i}^{y} e^{-2 \sum_{\bk, l=\cuof{1,2}} W_{\diamondsuit j_{l}}  \frac{g}{\omega_{A \bk  } \sqrt{N}}  \of{ a_{\bk}^{\dagger} e^{-i \bk \cdot \mathbf{r}_{j_{l}} } - a_{\bk} e^{i \bk \cdot \mathbf{r}_{j_{l}} }  } } \nonumber \\
\quad
\end{eqnarray}
Here, the two values of $j_{l}$ correspond to the locations of the two plaquettes which share the flipped spin. The action of $\sigma_{i}^{y}$ itself is trivial: it merely flips the sign of the two plaquettes $W_{\diamondsuit j}$ which share that spin. The expansion of the exponential factor, however, is more complex, and we will now treat it in some detail.

We first note that the value of the exponentiated operator depends critically on the values of $W_{\diamondsuit j_{1}}$ and $W_{\diamondsuit j_{2}}$, the two plaquette operators which share the flipped spin. If $W_{\diamondsuit j_{1}} = W_{\diamondsuit j_{2}}$, then the process will create or annihilate anyons, but if $W_{\diamondsuit j_{1}} = - W_{\diamondsuit j_{2}}$, then the process will simply move a pre-existing anyon from one plaquette to the other. With minor variations, this is identical to the ``sudden switching" problem of phonons interacting with an electronic bound state \cite{mahan}. We are principally interested in the probability of the exponentiated operator creating \textit{zero} propagating bosons; recalling that $s = \sum_\bk s_\bk$ and $\sqof{s_\bk , s_\bp}=0$, we can quickly calculate:
\begin{eqnarray}\label{Ldef1}
e^{-\lambda} & \equiv & \bra{0}  e^{-2 \sum_{\bk, l=\cuof{1,2}} W_{\diamondsuit j_{l}}  \frac{g}{\omega_{A \bk  } \sqrt{N}}  \of{ a_{\bk}^{\dagger} e^{-i \bk \cdot \mathbf{r}_{j_{l}} } - a_{\bk} e^{i \bk \cdot \mathbf{r}_{j_{l}} }  } } \ket{0}; \nonumber \\
\lambda & = & \sum_{\bk} f_\bk = \sum_\bk \frac{4 g^{2}}{\omega_{\bk}^{2}} \of{1 \pm \cos \bk \cdot \mathbf{d} },
\end{eqnarray}
where + corresponds to creating or annihilating anyons, - to moving them, and $\mathbf{d}$ is a unit vector joining the neighboring plaquette centers. This factor will renormalize the squared matrix elements of the shadow lattice couplings we will introduce shortly, and if we choose the shadow lattice couplings to have low enough energies, they will be unable to create or annihilate propagating bosons and we need only consider the renormalization factor $e^{-\lambda}$ when taking the bosons' dynamics into account. 

On the other hand, primary quantum errors, which have an energy-independent rate $\Gamma_P$, can create or destroy bosonic excitations in the resonator lattices, as there is no resonance condition which suppresses the boson-generating terms in Eq.~(\ref{acty}). We ultimately want to be able to ignore the dynamics of the bosons in our error correction calculation, since without propagating modes, the system can be simulated completely classically and the energies of spin flip events are trivial to predict. To do so, we will give the bosons intrinsic loss rates $\Gamma_{A/B}$, and given an average spin flip rate $\Gamma_P$, we can consider the propagating boson lattice to be empty of excitations whenever $\Gamma_{A/B} \gg \avg{g/\omega_\bk}_\bk \Gamma_P$. It is important to note that boson losses are harmless from a computational point of view; losing a boson of momentum $\bk$ has the result:
\begin{eqnarray}
e^{-s_{\bk}} a_\bk e^{s_{\bk}} &=& \nonumber a_\bk + \sum_j \frac{W_{Pj} g}{\omega_{A\bk} \sqrt{N}} e^{-i \bk \cdot r_{j}}.
\end{eqnarray}
The $W_P$ operators measure stabilizers, but as the different states in a topologically degenerate manifold cannot be distinguished by any combination of stabilizer measurements, their action is irrelevant. Further, so long as $\Gamma_{A/B} \ll -\mu_{A/B}$, the structure of the interaction potential should be unaffected by the bosonic loss rate. We can therefore ignore the effect of boson losses, except to assume that the propagating boson lattice remains in its ground state at nearly all times. In this limit, we only need to consider $H_c$ in Eq.~(\ref{HC}), such that the system can be simulated classically. We note that we choose to work in this limit primarily for reasons of convenience, as being able to ignore the propagating bosons makes numerical simulations dramatically simpler. If boson creation and annihilation is significant, we expect that the basic behaviors observed in the results section will be unchanged, though the resulting energy shifts may make passive anyon removal somewhat less efficient.

\begin{figure}
\includegraphics[width=3.25in]{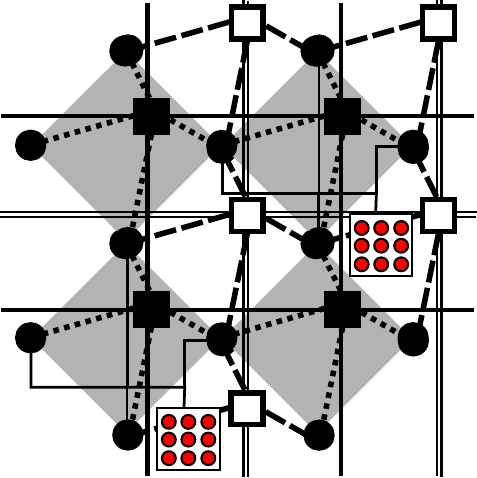}
\caption{(Color online) Primary lattice from FIG.~\ref{latfig} with the passive error correction module (shadow lattice) added. Each qubit (black circles) in the primary lattice is coupled to a group of shadow lattice qubits (the nine boxed red circles) which have rapid decay rates that shrink or eliminate excitations in the primary lattice created by unwanted interactions with the environment. A group of shadow lattice qubits may be shared between multiple primary qubits, and the couplings shown here are merely an example implementation where each group of nine shadow lattice qubits is shared by four primary lattice qubits. The shadow lattice qubits within each group are tuned so that each one has a distinct excitation energy, with the energies chosen to match transition energies in the primary lattice. This resonance condition leads to rapid error correction, while simultaneously minimizing additional errors generated by the shadow lattice itself.}\label{TCS}
\end{figure}

We now consider the effect of coupling this system to a shadow lattice, as shown in FIG.~\ref{TCS}. In this case, we will take the shadow lattice coupling to be of the form:
\begin{eqnarray}\label{HPS}
H_{PS} + H_S &=& \sum_{i,mn} \Omega_{mn0} \of{\sigma_{i}^{x} \sigma_{mnS}^{x}  + \sigma_{i}^{y} \sigma_{mnS}^{y} } \\ & & + \sum_{mn} \frac{\omega_{mn}}{2} \sigma_{mnS}^{z}, \nonumber
\end{eqnarray}
where the sites $i$ are spins on the primary (toric code) lattice and the $mn$ sites are spins on the shadow lattice. We choose a double index to denote shadow lattice sites for reasons that will quickly become apparent below; in computing transition rates induced by the shadow lattice, it is convenient to group the shadow lattice qubits by shared (and comparatively rapid) decay rates $\Gamma_{Sm}$, with the label $n$ denoting distinct qubit energies and couplings within the group $m$. All primary lattice sites couple to identical collections of shadow lattice sites; how those terms are distributed will be described below. The shadow lattice is a set of independent spins (or resonators), which are chosen to have fast decay rates and single-site energies which are resonant with excitations in the primary lattice. As we remarked earlier, we expect that the underlying quantum device architecture will be driven and studied in a rotating frame, so the shadow lattice energies will represent detunings from the primary qubit frequencies. The coupling between the two lattices causes excitations to tunnel back and forth from the primary lattice to the shadow lattice, but when the excitation is on the shadow lattice it decays rapidly, removing energy from the combined system and relaxing the primary lattice back toward its (rotating frame) ground state.

The shadow lattice can be easily integrated out through a Fermi's golden rule calculation, yielding a set of transition rates $\Gamma_{Ri}^{\alpha \beta}$ for the primary lattice to incoherently transition from an initial state $\ket{\alpha}$ to a final state $\ket{\beta}$ through a spin flip at site $i$. We first consider the coupling of a spin at site $i$ to a collection $m$ of distinct shadow lattice qubits (labelled by $n$), with energies $\omega_{mn}$ and a uniform decay rate $\Gamma_{Sm}$. The couplings between the two lattices have energy $\Omega_{mn0}$, and upon integrating out the shadow lattice, we arrive at the induced transition rate $\Gamma_{Rim}^{\alpha \beta}$:
\begin{eqnarray}\label{gammaRi}
\Gamma_{Rim}^{\alpha \beta} & = & \frac{\Gamma_{PSim}^{\alpha \beta} \Gamma_{Sm} }{\Gamma_{PSim}^{\alpha \beta} + \Gamma_{Sm} } ; \\
\Gamma_{PSim}^{\alpha \beta} & \equiv & \sum_{n} \frac{ M_{\alpha \beta i mn}^{2} \Gamma_{Sm} }{ \of{ \epsilon_{\beta} - \epsilon_{\alpha} + \omega_{mn}  }^{2} + \Gamma_{Sm}^{2}/4   }, \nonumber \\
M_{\alpha \beta i mn}^{2} & = & \Omega_{mn0}^{2} \abs{\bra{\beta} \sigma_{i}^{-} \ket{\alpha}}^{2}. \nonumber
\end{eqnarray}
Since the shadow lattice is coupled to single spins, $\ket{\alpha}$ and $\ket{\beta}$ must be related by a single spin flip for $\Gamma_{Rim}^{\alpha \beta}$ to be nonzero. Similarly, if a spin $i$ on the primary lattice is coupled to multiple groups $m$ of shadow lattice qubits with distinct decay rates $\Gamma_{Sm}$ for each group (as it will be in our simulations), then the total transition rate $\Gamma_{Ri}^{\alpha \beta}$ can be computed by simple summation ($\Gamma_{Ri}^{\alpha \beta} = \sum_m \Gamma_{Rim}^{\alpha \beta}$), provided that the energy ranges covered by the sets of shadow lattice qubits are well separated from each other in comparison to $\Gamma_{Sm}$ and the couplings $\Omega$. 

The transition rate between two states due to a single spin flip thus depends primarily on the energy difference between the two states. The fact that the shadow lattice couplings are comparatively weak enforces a resonance condition: to obtain an appreciable transition rate, there must be at least one shadow lattice qubit where the squared detuning $\of{\epsilon_{\beta} - \epsilon_{\alpha} + \omega_{mn}}^{2}$ is not significantly greater than the coupling and decay rates $\Gamma_{Sm}^{2}, \Omega_{mn}^{2}$. The transition rate is thus sharply peaked at $\epsilon_\beta - \epsilon_\alpha + \omega_{mn} = 0$, and this condition is vital to ensuring that the shadow lattice efficiently corrects errors without significantly inducing errors itself. For the device parameters proposed in this work, the resonance condition implies that no bosons are created, since the bosons themselves are fairly high energy objects and will cause the process to be off resonance and thus rare. This stands in contrast to the primary error rate, which we assume is constant across the lattice and independent of the energetics of the many-body system. Recalling that the probability of a given spin flip creating zero bosons in this system is $e^{-\lambda}$, where $\lambda = \sum_\bk f_\bk$ and depends critically on whether anyons are being created, annihilated or simply moved, the transition amplitude where no propagating bosons are created is
\begin{eqnarray}\label{defM}
M_{\alpha \beta i mn}^{2} = e^{- 2\lambda} \Omega_{mn0}^{2} \equiv \Omega_{mn}^{2}.
\end{eqnarray}
Here, we have used the near-resonance condition to merge the $\alpha$ and $\beta$ dependence of $M$ into $\Omega_{np}$. The two possible values of $\lambda$ (arising from moving or creating/annihilating anyons) correspond to processes on very different energy scales, and these processes will only occur at appreciable rates through coupling to shadow lattice sites where $\omega_{np} \simeq \epsilon_{\alpha} - \epsilon_{\beta}$. Neglecting the off resonant creation and annihilation of bosons, we drop the $\alpha$ and $\beta$ indicies and arrive at the final expression:
\begin{eqnarray}\label{GRfinal}
\Gamma_{Ri} \of{\delta E_{i}} &=& \sum_{m} \frac{ \Gamma_{PSm} \of{\delta E_i} \times \Gamma_{Sm} }{\Gamma_{PSm} \of{\delta E_i} + \Gamma_{Sm}  }; \\
\Gamma_{PSm} \of{\delta E_i} & = & \sum_n \frac{ \Omega_{mn}^{2} \Gamma_{Sm} }{ \of{ \delta E_i + \omega_{mn}}^{2} + \Gamma_{Sm}^{2}/4  }. \nonumber
\end{eqnarray}

We will refer to $\Gamma_{Ri} \of{\delta E_i}$ as the \textit{repair function} of the lattice, and it is the primary result of this subsection. The energy $\delta E_i$ is the energy change in the primary lattice from a spin flip at site $i$, and we employed this expression in all of our numerical simulations to estimate the transition rate induced by the shadow lattice. Each spin $i$ on the primary lattice will randomly flip with rate $\Gamma_P + \Gamma_{Ri}$, where $\Gamma_P$ is independent of the system's energetics but $\Gamma_{Ri}$ is not. Thanks to the commutivity of all operators in the limit where the coupling to the shadow lattice is approximated by a set of incoherent transitions at rates given by the repair function (\ref{GRfinal}), the system's time evolution occurs only through incoherent processes and can be simulated classically. The primary lattice has a degenerate ground state (with the degeneracy set by boundary conditions), and quantum information can be encoded in this degeneracy. States within the system's degenerate manifold will be mixed when an anyon pair (pair of $Q = 1$ values) is created, separates so that the anyons make a complete circuit of the system (e.g. wrap around one of the axes of a torus) before rejoining each other and being annihilated.  To study this system as a quantum memory for a given set of parameters, then, we initialize it in a state where all $Q$ operators evaluate to 0, and then evolve time through random spin flips until a created pair of anyons makes a complete circuit of the system and annihilates leaving a non-contractible loop. By repeating such a simulation many times we will arrive at a mean time to failure $\tau$, which is the lifetime of a quantum state encoded in the topological degeneracy of our system.

\begin{figure}
\includegraphics[width=3.0in]{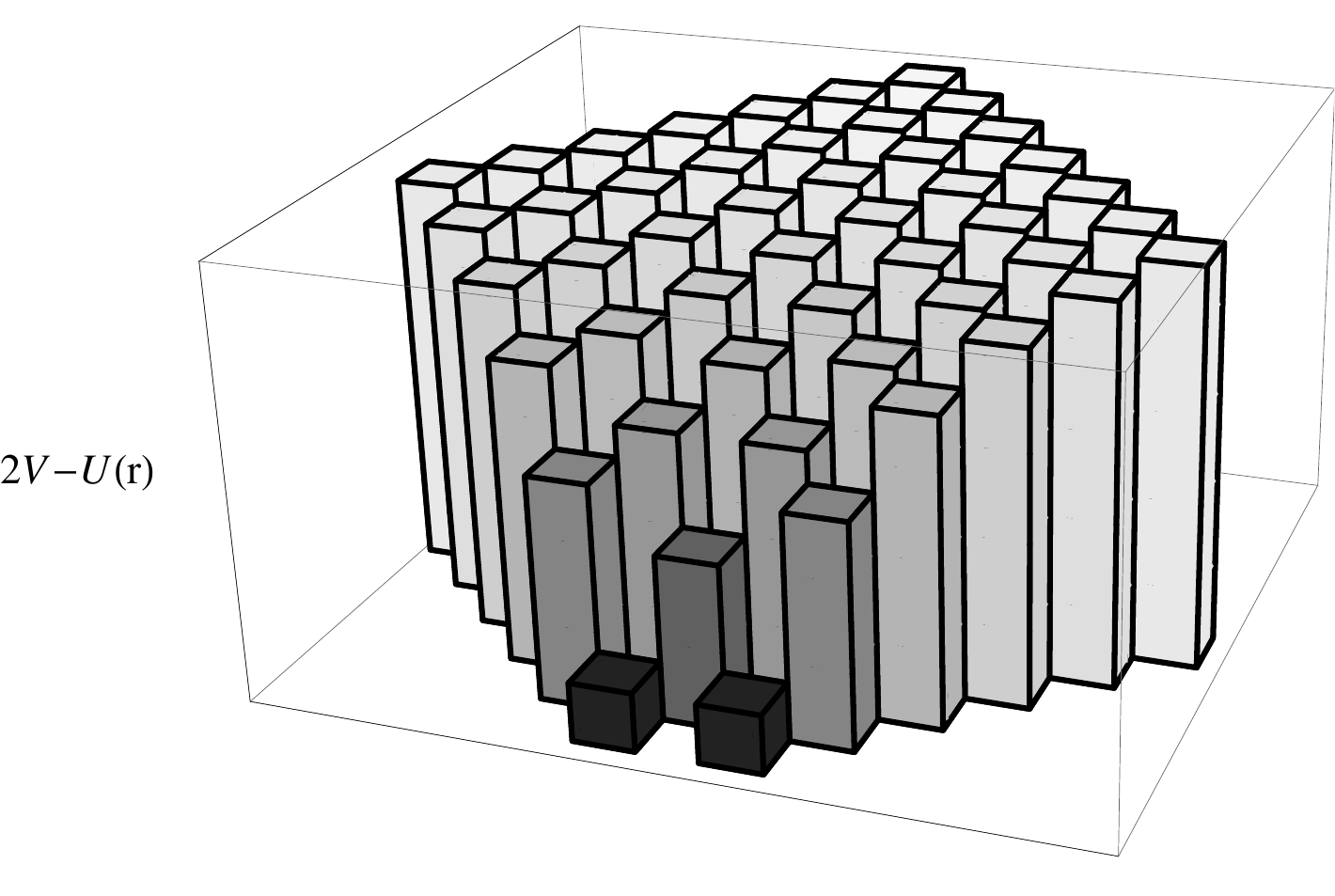}
\caption{Potential landscape for an anyon pair. The plaquette at $\mathbf{r}=0$ is not drawn, and is assumed to contain a single anyon; the heights of all other plaquettes correspond to the total energy $2V-U\of{r}$ of placing a second anyon on that plaquette. Note that as $U \ll V$ the minimum energies on the plot are still large. A spin sits at each shared corner, and by flipping the spin anyons can be created, annihilated, or exchanged between plaquettes. For each flip which brings anyons closer together, the system loses energy, and with a properly tuned shadow lattice, these flips will occur rapidly at a rate $\Gamma_{R} \of{\delta E}$. As spin flips which move anyons further apart occur at a much slower primary error rate $\Gamma_P$, the probability for a local pair of anyons to separate widely after creation is suppressed by a factor $ \propto \prod_{n = 1}^{n_{max}} \of{ \Gamma_R \of{\delta E_n } / \Gamma_P  } $, where $n_{max}$ is the average maximum number of steps it would take to separate anyons such that the potential differences for further steps are too small for the shadow lattice to act upon. For $\Gamma_R \gg \Gamma_P$, this can lead to an enormous increase in the lifetime of quantum information encoded in the system's topological ground state degeneracy.}\label{Urfig}
\end{figure}

\begin{figure}
\includegraphics[width=2.25in]{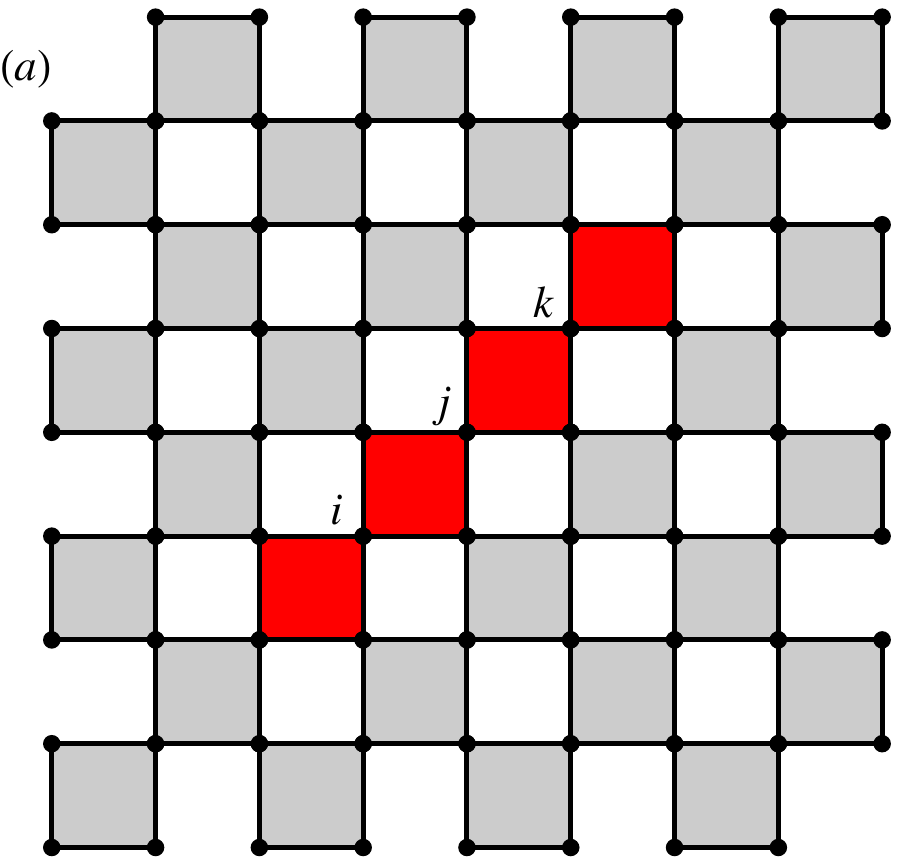}
\includegraphics[width=2.25in]{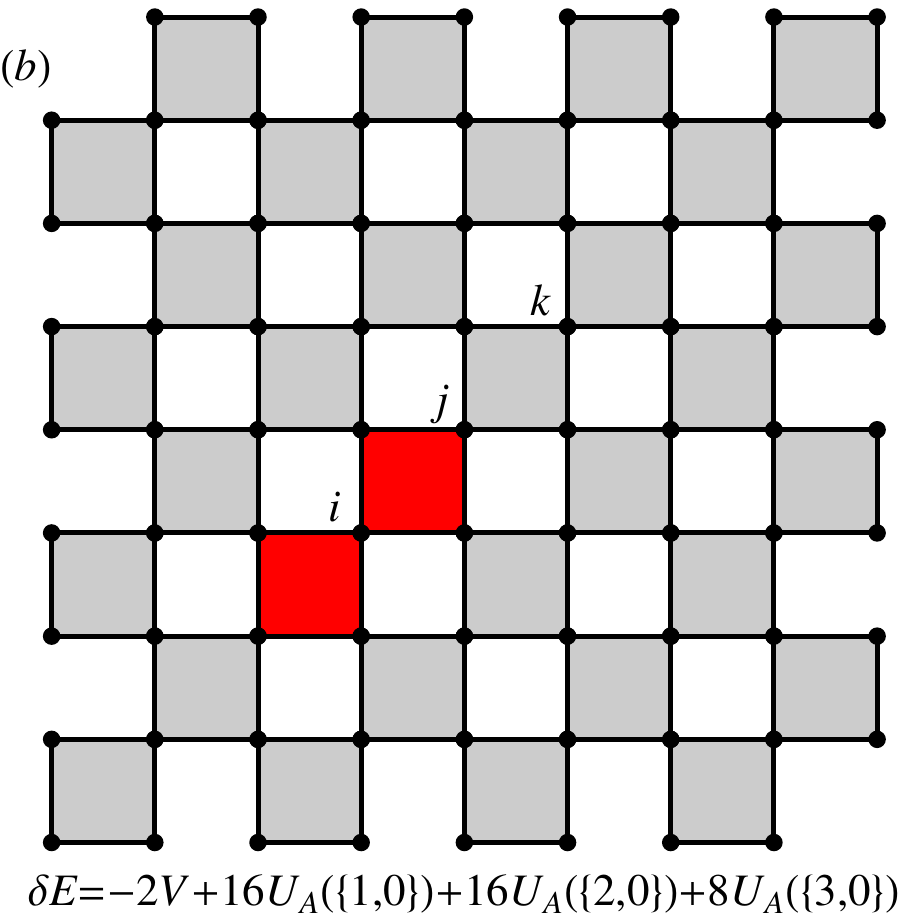}
\includegraphics[width=2.25in]{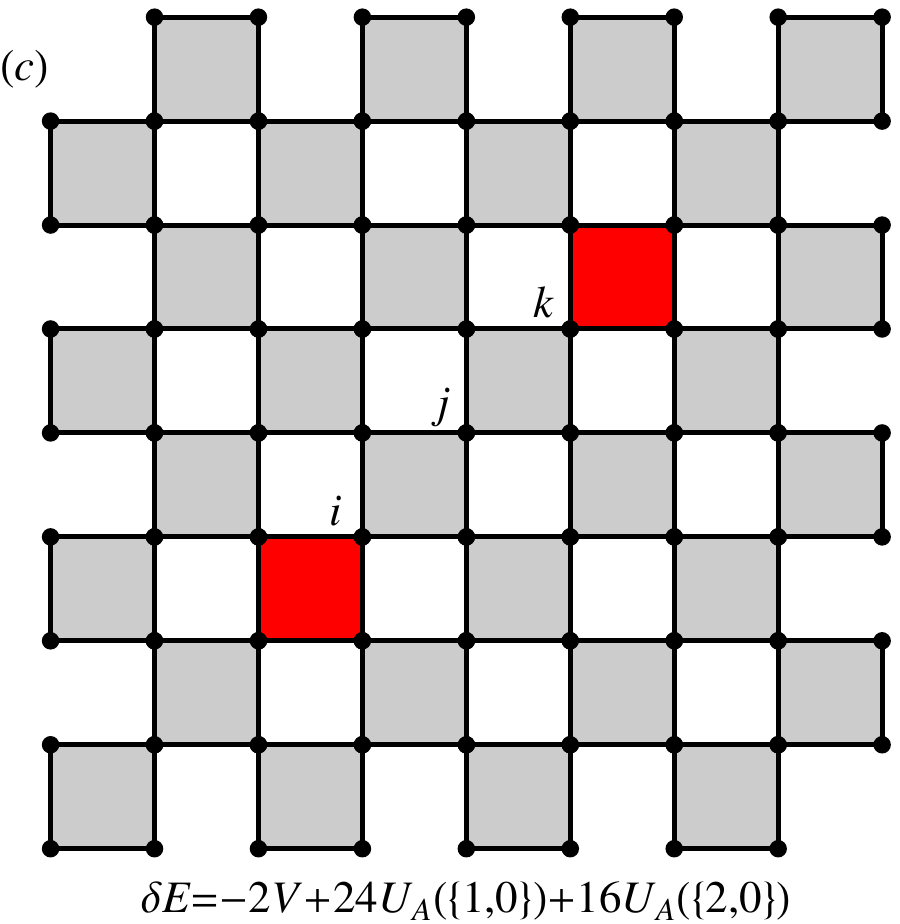}
\caption{(Color online) Anyon removal as an error source. In (a), spins $i$ and $k$ have flipped, creating four anyons in the adjoining plaquettes. The shadow lattice will rapidly correct these errors, and if spin $i$ or $k$ flips as shown in (b), the remaining anyon pair can be quickly eliminated, returning the system to its ground state. If however spin $j$ flips first as in (c), the remaining anyon pair will require three further operations to eliminate and is thus much more likely to lead to a quantum logic error than in case (b). To reduce the likelyhood of this process, we add a second set of shadow lattice qubits which are tuned to be close to resonance with the energy removed in process (b) but not process (c), ensuring that process (b) will occur much more rapidly, making process (c) rare.}\label{targetfix}
\end{figure}

\subsection{Tuning the shadow lattice}

We will now state loose guidelines for choosing shadow lattice parameters to maximize $\tau$ for a given $\Gamma_P$, $V_{A/B}$ and bosonic dispersion. Our goal here is not to identify the \textit{best} possible shadow lattice, as the shadow lattice's design in a real system would be constrained by many experimental factors which are beyond the scope of this paper, but merely to identify a \textit{good} set of parameters, and to identify general constraints and scaling behaviors. Throughout, we will assume that $V_{A/B} \gg g$, so that the energy scale for creating or annihilating anyons is well-separated from the energy scale for simply moving them relative to each other.

We first consider the resonant removal of local anyon pairs, the most basic form of error correction in our system. To do this, we couple each site or collection of sites to a shadow lattice qubit as in (\ref{HPS}); all shadow lattice couplings will be exchange couplings of the form $\sigma_{i}^{x}\sigma_{iS}^{x} + \sigma_{i}^{y} \sigma_{iS}^{y}$. We will refer to these qubits as the first primary repair qubits, and choose their energy $\omega_1 = 2 V_{A/B} - 8 U_{A/B} \of{\mathbf{d}}$ to be resonant with the excitation energy of a pair of defects plaquettes sharing a single spin \footnote{To efficiently correct both $x$ and $y$ errors that arise from photon losses ($\sigma^-$ operator) using simple transfer couplings, one should choose $V_A \neq V_B$, with $\abs{V_A - V_B} \gg \Omega$, necessitating two sets of primary repair qubits. The physics of the two types of error correction is otherwise identical, so we only treat one type of quantum error in this writeup. More complex coupling types could allow for us to set $V_A = V_B$ and use only one set of primary repair qubits to remove both plaquette and star anyons. }. We choose the coupling $\Omega_1$ and loss rate $\Gamma_{S1}$ to both be small compared to $\omega_1$, so that the flip rate induced by this component of the shadow lattice is negligible for any processes which do not correspond to repairing local pairs of defects. Steps which create or separate anyons occur due to the primary error rate (photon losses into the outside world, in a circuit QED system), and thus have an energy-independent rate $\Gamma_P$. Provided $\omega_1$ is sufficiently large, $\Omega_1$ and $\Gamma_{S1}$ can still both be large compared to $\Gamma_P$, so the shadow lattice is able to rapidly eliminate local anyon pairs when they are created. However, as soon as defects become spatially separated (so that their corresponding plaquettes do not share a spin), this component of shadow lattice can no longer help, as defect removal is a purely local process. The lifetime of the device in this limit therefore scales as $\Gamma_R \of{-2V}/ \Gamma_{P}^{2}$, but is expected to remain flat or slowly decrease as the system size increases.

Before considering anyon motion, we wish to point out that the quantum state lifetime of the device can be further improved by adding second set of shadow lattice qubits with energies near $2V_{A/B}$. The reason for doing so is shown in figure FIG.~\ref{targetfix} Consider a pair of quantum errors, separated so that four plaquettes are flipped along a connected chain. In this case, there are three spins which can flip to remove two anyons, all of which can be acted on by the shadow lattice. If either the first or last spin flips, then there is no problem, as the pair of anyons which remain are local and will themselves be quickly removed by a second shadow lattice term. However, if the central spin flips, the remaining pair will be separated and require three spin flips to eliminate, making them much more likely to wander apart and eventually cause a logical error. These processes are particularly dangerous to the encoded quantum information, but thanks to the ranged interaction, these processes remove different amounts of energy, and so to reduce the probability of flipping the central spin first, we add a second set of shadow lattice qubits (called the second primary repair qubits) with a somewhat narrower resonance (smaller $\Omega$ and $\Gamma_S$ than the first primary repair qubits), tuned to match the energy of flipping either of the outermost spins. This makes it much more likely that the first or last spins will flip before the center spin does, significantly mitigating this error source. In our simulations, we found that this second set of primary repair qubits can increase the average lifetime by a factor of 2 or more, so we included them in all the calculations described below.

Of course, even with primary repair qubits, multiple spin flips can still lead to separated anyons, which will wander and can eventually cause logical errors. Correcting this type of error requires ranged interactions, but once these interactions are introduced, a small number of shadow lattice qubits can lead to efficient, multistage passive error correction. Each step that further separates a pair of anyons increases the energy of the many-body system by $\delta E $, which depends on the relative positions of the anyons, and if the shadow lattice is tuned to remove energy (through a collection of \textit{secondary repair} qubits) in this range at some rate $\Gamma_R \of{\delta E}$, then the relative likelyhood of a pair of separated anyons moving one step further apart compared to moving back together is suppressed by a factor $\propto \Gamma_R \of{\delta E}/\Gamma_P$. The total lifetime can therefore be increased by a factor  $ \propto \prod_{n = 1}^{n_{max}} \of{ \Gamma_R \of{\delta E_n } / \Gamma_P  } $, where $n_{max}$ is the average number of steps to separate anyons such that the potential is no longer strong enough to make the cost of separating them further significant. For appropriate device parameters, $n_{max}$ can be quite large, meaning that reducing the primary error rate $\Gamma_P$ by a factor of $a$ can reduce the logical error rate by $\sim a^{n_{max}}$, leading to an enormous increase in the quantum state lifetime, and thus, efficient, passive, quantum logical error correction. 

However, such dramatic improvements are only observed below a threshold value of $\Gamma_P$, which depends on the system size and the gap parameter $\mu$ that controls the range and strength of the potential. Essentially, fluctuating pairs of anyons that incoherently appear and vanish will exert random forces on isolated anyons, potentially dragging them further apart. Further, nearby pairs also shift the local energy, pushing the primary repair qubits off resonance and reducing the repair rate. The significance of these effects grows with both the strength and the range of the potential, so depending on $\Gamma_P$ and the size of the lattice, decreasing $\mu$ may make the overall lifetime shorter. Similarly, if the lattice reaches a large enough size (compared to the range $\sqrt{-J/\mu}$), further increases in lattice size will begin to reduce the quantum state lifetime, as each additional spin is another potential error source to create anyon pairs.

\begin{figure}
\includegraphics[width=3.25in]{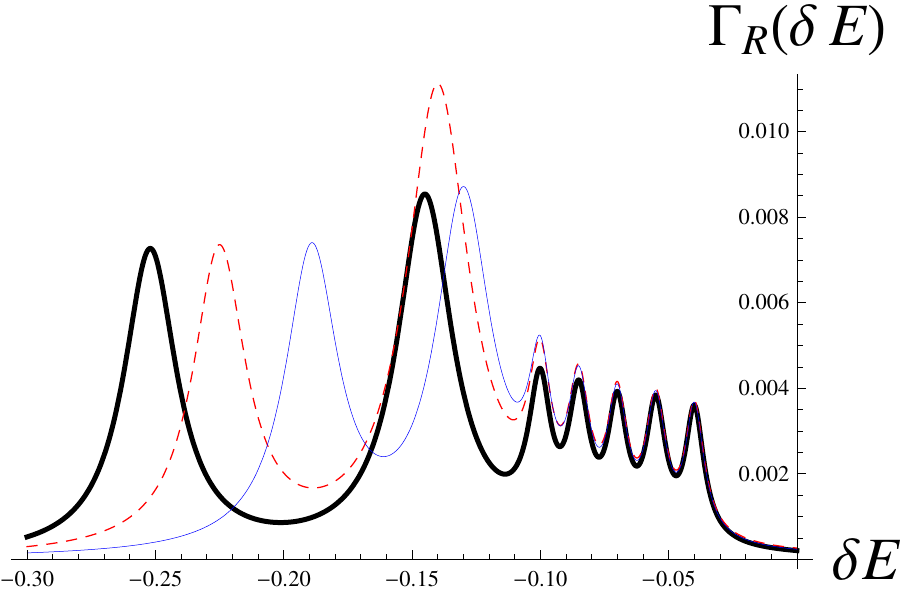}
\caption{(Color online) Repair functions $\Gamma_{R} \of{\delta E}$ used in our simulations. The three curves plot the repair rates generated by shadow lattices tuned to pull separated anyons back together for $-\mu_A = 0.1 J$ (thick, black), $0.2 J$ (dashed, red) and $0.4 J$ (thin, blue), with $J=3g$ and all energies expressed in units of $g$. The primary repair terms which annihilate anyons are targeted at energies $\delta E \sim - 60 g$, so have negligible influence on processes in the region plotted. The precise parameters of each shadow lattice are detailed in the appendix.}\label{repairfunction}
\end{figure}

In our numerical simulations, we let each spin in the primary lattice be coupled to two primary and seven secondary repair qubits, yielding the repair functions shown in FIG.~\ref{repairfunction}. The seven secondary repair qubits have energies and couplings tuned to match common stepwise changes in the system's energy from separating nearby anyons. The two highest energy secondary repair qubits have energies which were well-separated from each other; the five lowest energy secondary repair qubits are spaced much more closely, to cover the range of lower energy transitions that occur when further separating anyons that are already a few plaquettes apart. By covering a large range of transition energies, the shadow lattice causes nearby anyons to fall back together much more quickly than they can separate, leading to potentially enormous increases in the device's quantum state lifetime. 

We note finally that there is a great deal of arbitrariness in the parameter choices studied here. First, as our simulations computed mean time to failure, an exponential increase in device lifetime meant an exponential increase in simulation runtime-- in other words, the more effective the error correction, the more difficult it was for us to prove that effectiveness. Due to the computational complexity, we did not attempt to numerically optimize the shadow lattice, and instead chose parameters ``by hand," through the ad hoc prescription of attempting to maximize the range of energies covered and correction rate $\Gamma_{Ri} \of{\delta E_i < 0}$ while minimizing the additional error rate induced by the shadow lattice ($\Gamma_{Ri} \of{\delta E_i >0}$). All that said, we will now present the results of numerical simulations (with the shadow lattice taken as a spatially uniform, energy dependent rate of repair $\Gamma_{Ri} \of{\delta E_i}$), and demonstrate passive quantum logical error correction in a relatively realistic qubit model.

\begin{figure}
\includegraphics[width=3in]{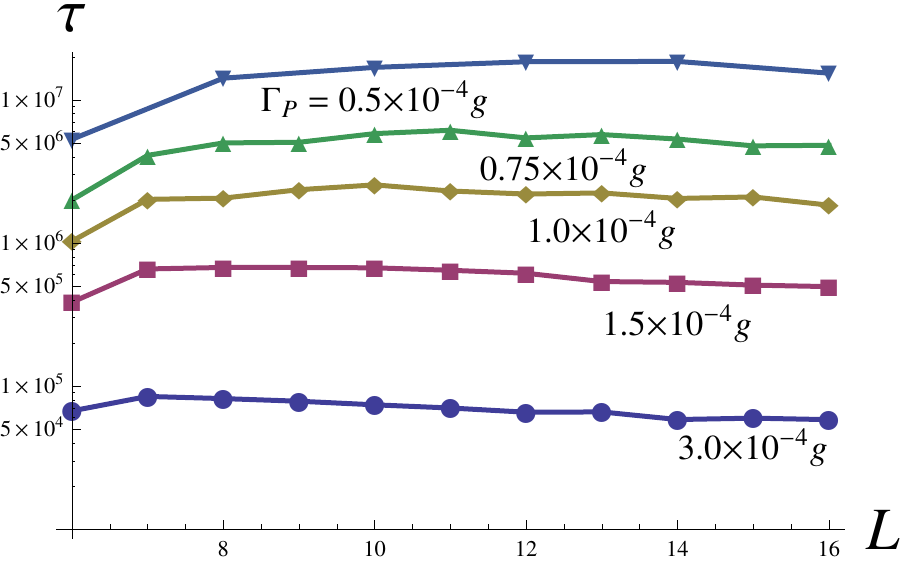}
\includegraphics[width=3in]{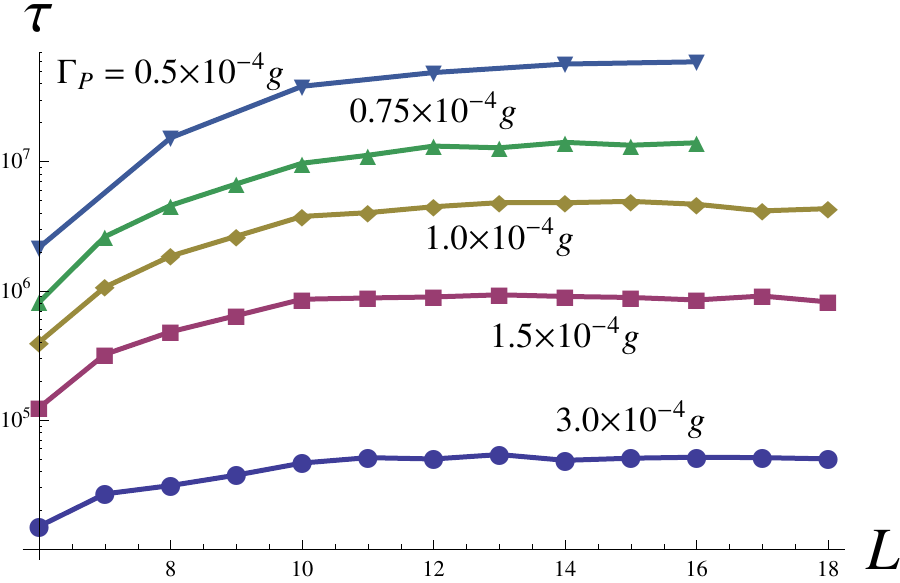}
\includegraphics[width=3in]{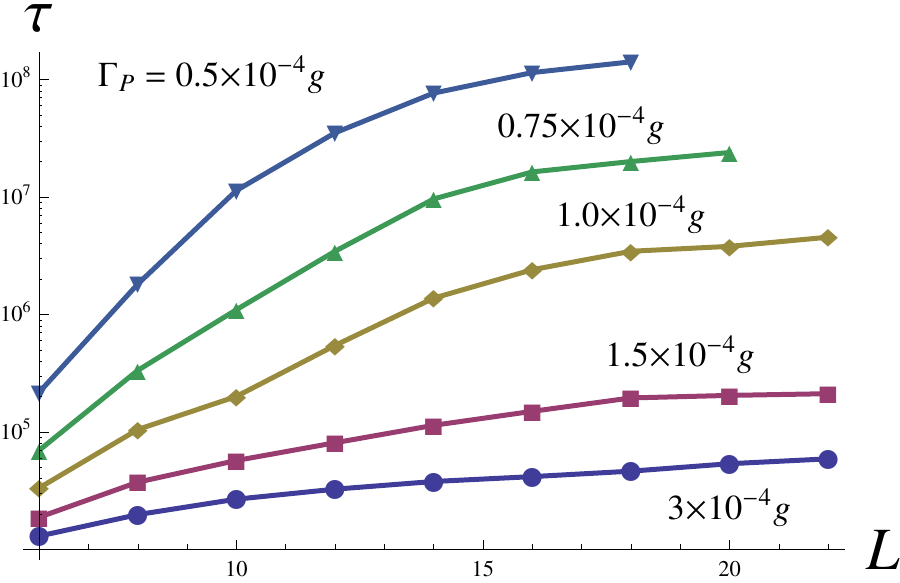}
\caption{(Color online) Quantum state lifetime $\tau$ (in units of $g^{-1}$, where $g$ is the driven plaquette/star coupling to the resonator array) of a logical state encoded in the topological quantum degeneracy of an $L \times L$ toroidal implementation of our qubit array. From top to bottom, the systems simulated have propagating boson gaps $-\mu/J = 0.4$, 0.2 and 0.1; the shadow lattice parameters used for passive error correction are detailed in Table I. In each plot we display the average lifetime $\tau$ computed for (bottom to top curves) primary error rate $\Gamma_P = \cuof{ 3, 1.5, 1, 0.75, 0.5} \times 10^{-4} g$. The best and worst average lifetimes come from systems which differ by mere factors of $3$ in $L$ and 6 in $\Gamma_P$, but over $10^{4}$ in $\tau$ itself. Note also that the best case studied ($18 \times 18$, $-\mu/J = 0.1$, $\Gamma_P = 0.5 \times 10^{-4} g$) has an average lifetime around seven thousand times longer than any of the individual qubits. Due to the finite length scale $l = \sqrt{-J/\mu}$ of the interaction potential, the error correction in each case saturates around $L_c \approx 7 l $. Beyond $L = L_c$ increasing the size of the lattice simply adds more potential anyon sources without significantly increasing the probability of anyons being recaptured once they reach the random walk regime (a separation of more than $\sim 2 l$ for the parameters considered). Each data point for $\Gamma_P = 0.5 \times 10^{-4} g$ was generated from 400 error simulations.}\label{datafig}
\end{figure}

\section{Results}\label{ressec}

\begin{table}
\caption{Shadow lattice parameters used in our numerical simulations. The values of $\Omega$ implicitly include the modifying factors $e^{-\lambda}$ enforced by the requirement that no propagating bosons are created (Eq.~\ref{defM}). The shadow lattice qubits are divided into four groups (qubit 1; qubit 2; qubits 3 and 4; qubits 5-9) based on their relaxation rates $\Gamma_{S}$, and repair quantum errors at rates computed from the repair function (\ref{GRfinal}). All terms are chosen to match transitions with $V_A = 30 g$ and $J = 3g$ and are in units of $g$. \label{pars}}
\begin{ruledtabular}
\begin{tabular}{|c|ccc|}
$-\mu/J$ & 0.1& 0.2 & 0.4 \\
\hline
$ \omega_1 $ & 59.33 & 59.55 & 59.67 \\
$  \Omega_1, \Gamma_{S1} $ & $ \cuof{ 0.33, 0.33 } $ & $ \cuof{ 0.33, 0.33 } $ & $ \cuof{ 0.33, 0.33 } $ \\
\hline
$\omega_2$ & 58.13 & 58.6 & 59.07 \\
$   \Omega_2, \Gamma_{S2} $ & $ \cuof{ 0.267, 0.2 } $ & $ \cuof{ 0.267, 0.2 } $ & $ \cuof{ 0.267, 0.2 } $ \\
\hline
$\omega_3$ & 0.252 & 0.225 & 0.189 \\
$  \Omega_3, \Gamma_{S3}  $ & $\cuof{0.0075,0.02  } $ & $\cuof{0.0075,0.02 } $ & $\cuof{0.0075,0.02 } $ \\
\hline
$\omega_4 $ & 0.145 & 0.14 & 0.13 \\
$  \Omega_4, \Gamma_{S4}  $ & $\cuof{0.0085,0.02  } $ & $\cuof{0.011,0.02  } $ & $\cuof{0.0085,0.02  } $ \\
\hline
$\omega_5 $ & 0.1 & 0.1 & 0.1 \\
$ \Omega_5, \Gamma_{S5}  $ & $\cuof{0.005,0.0045  } $ & $\cuof{0.005,0.0045  } $ & $\cuof{0.005,0.0045  } $ \\
\hline
$\omega_6$ & 0.085 & 0.085 & 0.085 \\
$  \Omega_6, \Gamma_{S6}  $ & $\cuof{0.005,0.0045  } $ & $\cuof{0.005,0.0045  } $ & $\cuof{0.005,0.0045  } $ \\
\hline
$\omega_7$ & 0.07 & 0.07 & 0.07 \\
$ \Omega_7, \Gamma_{S7}  $ & $\cuof{0.0045,0.0045  } $ & $\cuof{0.0045,0.0045  } $ & $\cuof{0.0045,0.0045  } $ \\
\hline
$ \omega_8$ & 0.055 & 0.055 & 0.055 \\
$ \Omega_8, \Gamma_{S8}  $ & $\cuof{0.0045,0.0045  } $ & $\cuof{0.0045,0.0045  } $ & $\cuof{0.0045,0.0045  } $ \\
\hline
$\omega_9$ & 0.04 & 0.04 & 0.04 \\
$ \Omega_9, \Gamma_{S9}  $ & $\cuof{0.004,0.0045  } $ & $\cuof{0.004,0.0045  } $ & $\cuof{0.004,0.0045  } $ 
\end{tabular}
\end{ruledtabular}
\end{table}

We numerically studied a variety of system sizes, error rates, and shadow lattice parameters, with the goal of demonstrating both exponential logical error suppression and its limits as the system grows. In all cases, we chose $J = 3 g$ and $V_{A} = 30 g$ in Eq.~(\ref{HC}), such that the energy barrier $\sim 2V_A$ the system must overcome to transition between ground states was nearly constant across all periodic simulations. As the system is coupled to an effectively infinite temperature bath due to loss processes, the precise value of the energy barrier is irrelevant so long as it is large compared to the motional energy scale $g^{2}/J$ and appropriately matched by the shadow lattice. In the simulations with physical edges, the minimum energy barrier was simply $V_A$, as single anyons can be created at either the top or bottom edge. Unless otherwise noted, each data point is the average of 900 individual runs of the defect tracking algorithm discussed in appendix A.

\subsection{Periodic boundary conditions}

The bulk of our numerical studies employed periodic boundary conditions, where the primary lattice forms a torus of $L \times L$ plaquettes. In this case, the ground state is actually four-fold degenerate, with degenerate states being mixed by a pair of anyons wrapping around either axis of the torus. For simplicity, we restricted ourselves to only simulating a single type of quantum error (random $\sigma^y$ operations which create plaquette defects) and only tracked the decoherence of a single logical qubit, so that errors created by anyons traversing a vertical path were tracked but errors created by horizontal anyon motion were not. Horizontal chains of $\sigma^x$ errors create a second error channel for our logical qubit, but as the physics of these errors and their correction is identical to vertical anyon motion, we ignored them. The bit flip and decoherence rates of our logical qubit are thus equal to the values we report below.

We simulated error propagation for lattice sizes ranging from $6 \times 6$ to $22 \times 22$ plaquettes, for three values of $-\mu$ (0.4, 0.2, and 0.1 $J$) and five values of the primary error rate $\Gamma_P$. As the interaction decays exponentially at long ranges, the system's sole length scale $l$ is the inverse of the exponential decay constant $\sqrt{-\mu/J}$, with $l = 1.58$, 2.24 and 3.16 for $-\mu = 0.4$, 0.2 and 0.1 $J$. The shadow lattice parameters were chosen to match the transition energies of separating anyons with an interaction $U \of{\mathbf{r}}$ given by Eq. (\ref{tildeH}), generating the repair functions shown in FIG.~\ref{repairfunction} and the quantum state lifetimes reported in FIG.~\ref{datafig}. If $\Gamma_P$ is small enough, many shadow lattice stages can efficiently pull anyons back together and eliminate them, leading to dramatic increases in the quantum state lifetime. For example, in the numerical simulations shown in FIG.~\ref{datafig}, for a $20 \times 20$ plaquette lattice with periodic boundary conditions and $\mu = -0.1 J$, reducing the primary error rate by a factor of two from $\Gamma_P = 1.5 \times 10^{-4} g$ to $7.5 \times 10^{-5} g$ increases the quantum state lifetime by a factor of 120, or nearly $2^7$. In this limit, an average of around 1.5 million random bit flips ($\Gamma_P$ processes) occurs between each logical error, starkly demonstrating the effectiveness of the passive error correction.

For the range of parameters studied the system's quantum state lifetime $\tau$ increases continuously with increasing lattice size $L$ until a saturation point at $L_c$ is reached. We observed that the value of $L_c$ depends somewhat on $\Gamma_P$ and tends to increase slowly as $\Gamma_P$ decreases. For $L > L_c$, $\tau$ ceases to improve and instead fluctuates or slowly decreases as $L$ is increased further. We attribute this effect to the finite range of the potential; once anyons are separated by more than $\abs{\mathbf{r}} \simeq 2 l$, the energy cost of separating them further falls outside the range where the shadow lattice is near a resonance. When this occurs the repair rate $\Gamma_{Ri}$ is not much greater than the primary error rate $\Gamma_P$, so anyons move in a random walk through further $\Gamma_P$ processes until they approach another anyon and fuse. As each lattice spin is another potential source of new anyon pairs, and additional sites do nothing to make anyons more likely to return to each other once they enter the random walk regime, adding more sites eventually becomes counterproductive and will slowly decrease the logical state lifetime.

\begin{figure}
\includegraphics[width=3.25in]{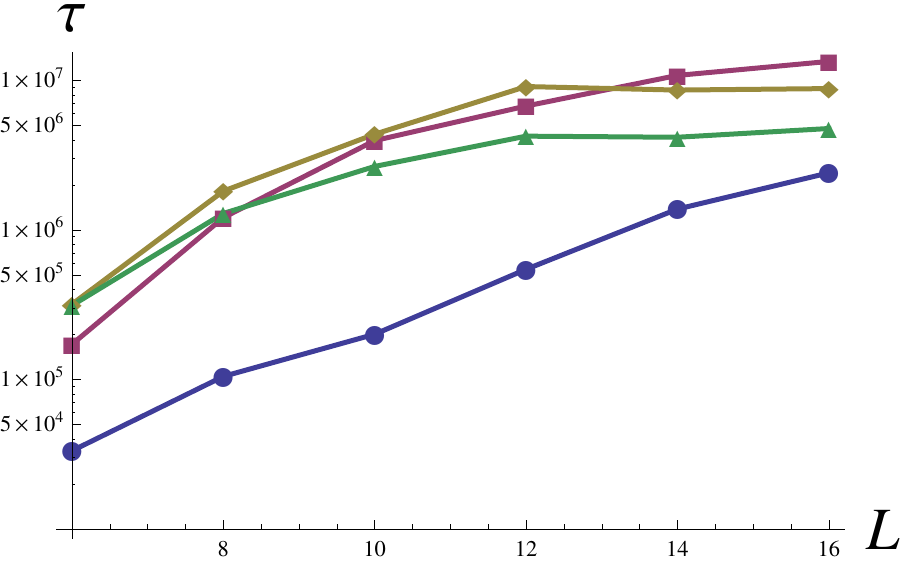}
\caption{(Color online) Comparison of different primary repair strengths for $\mu/J = -0.1$ and $\Gamma_P=10^{-4}$. The curves are computed using the secondary repair parameters listed in Table~\ref{pars} and varying primary repair strengths. We plot the results for the primary repair parameters in Table~\ref{pars} (blue, circles), a 50\% increase the values of $\Omega_1,\Omega_2,\Gamma_{S1},\Gamma_{S2}$ (purple, squares), a 100\% increase (gold, diamonds), and a 150\% increase (green, triangles). The optimal values of the primary repair terms depend on the size of the lattice, but making them too strong eventually becomes counterproductive and reduces the lifetime, as it reduces the selectivity of the error correction process discussed in FIG.~\ref{targetfix}. Each data point shown is the average of 400 error simulations.}\label{modP}
\end{figure}

Multi-anyon processes, where the interactions between four or more anyons push the system's energetics off resonance with the shadow lattice and inhibit error correction or drag separated anyons further apart, are extremely significant contributions to the net logical error rate, particularly at small $\mu$. We present evidence for this in FIG.~\ref{modP}, where we compare systems with identical primary error rates and secondary repair qubits, but different primary repair rates (removal of local pairs of anyons through the shadow lattice). If multi-anyon effects are insignificant, then doubling the primary repair rate while holding all other parameters fixed should improve the lifetime $\tau$ by roughly a factor of two, as once anyons are separated the primary repair qubits can do nothing to remove them unless they are pulled back together by secondary repair qubits. Conversely, if multi-anyon effects are important, the lifetime should improve by a larger factor, since the instantaneous density of fluctuating anyon pairs (and their subsequent influence on other anyons) will be reduced. Our numerical simulations confirm that multi-anyon effects are an extremely important contribution to the logical error rate, particularly for $-\mu = 0.1 J$; as shown in FIG.~\ref{modP}, doubling the primary repair rates can produce an improvement of more than a factor of twenty in the logical state lifetime for small lattices.

There are limits, of course, on how strong the primary repair couplings $\Omega$ and $\Gamma_S$ can be. If the coupling to the shadow lattice is strong, errors introduced by the shadow lattice itself will become significant, and the resulting perturbative corrections will eventually interfere with state readout in a real system. Further, making the two primary repair qubits too strongly coupled washes out the difference in energies that suppresses the local multi-step processes shown in FIG.~\ref{targetfix}, which can actually reduce the final lifetime by making this error channel stronger. For larger lattices, suppressing the local multi-step contribution is more important than further increasing the primary repair rate; as seen in FIG.~\ref{modP}, the optimal strengths of the first and second primary repair terms depend on $\mu$ and the dimensions of the lattice.

It worth pausing to compare the parameters used in our theoretical simulations to realistic decay and coupling rates in circuit QED systems. Recent progress in transmon \cite{kochyu,houckkoch,rigettipoletto2012,changvissers} and fluxonium qubits \cite{popgeerlings2014,voolpop2014} has yielded loss and decoherence rates as low as $\Gamma = 10$kHz, corresponding to state lifetimes of $100 \mu$s. If we take $\Gamma_P = 10$kHz as our global error rate, then to obtain $\Gamma_P = 10^{-4} g$ requires $g/\hbar \simeq 2 \pi \times 16$MHz, which is quite large for a coupling generated at high order in perturbation theory from a gadget construction, such as the one detailed in appendix B. Simultaneously achieving qubit nonlinearities large enough to accommodate strong perturbative couplings while maintaining low loss rates is a challenging task. That said, progress in circuit QED architectures has been extremely rapid in recent years, and we expect that both lifetimes and nonlinearities will be substantially improved in the next decade. For instance, loss rates below $1$kHz have been observed in fluxonium devices, though the dephasing rate ($\sigma^z$ errors) remains at $50$kHz or higher. If dephasing were brought in line with decay for these qubits, we would only need to achieve $g/\hbar = 2\pi \times 1.6$ MHz, which is much more feasible. We note also that we have assumed a white noise model for our simulations, appropriate to bit flips generated by photon losses. Dephasing in realistic superconducting devices often comes from power law noise and is dominated by low frequency processes, often at frequencies much less than the qubit-qubit interactions we consider here, and thus would have difficulty inducing transitions in many-body states. In comparison to measurement based schemes (where the qubits are uncoupled except through repeated pulse sequences), we expect that the physical Hamiltonian of our system could provide some level of additional protection against dephasing noise, though a detailed comparison of active and passive quantum error correction for various noise spectra is beyond the scope of this work.

\subsection{Physical edges}

\begin{figure}
\includegraphics[width=3.25in]{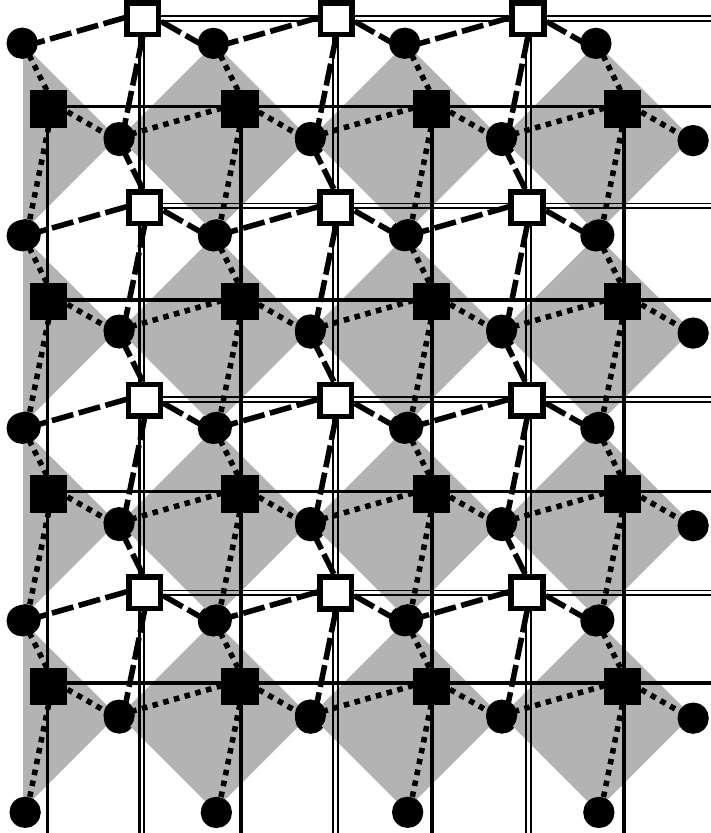}
\caption{Implementation of edges in our passive error correction scheme. As in the surface code \cite{fowlersurface}, non-periodic boundary conditions are implemented by truncating four-body operators to three body terms; along the top or bottom edges, stars are truncated, and along the left or right edges plaquettes are truncated. The figure depicts the upper left corner of a larger rectangular patch of quantum computing fabric.}\label{edgefig}
\end{figure}

While this work has so far only considered versions of our system with periodic boundary conditions, systems with physical edges are far easier to implement, and in the surface code framework, operations on logical states are frequently performed by exploiting the edge structure or even ``cutting holes" in the lattice and braiding the holes around each other. Edges in a square patch of surface code/toric code are implemented as shown in FIG.~\ref{edgefig}, where on the top and bottom edges, star terms are truncated into three-body interactions, and on the left and right edges plaquette terms are truncated. The system has a twofold degenerate ground state, which can be parametrized by the eigenvalue of a string operation consisting of a chain of $\sigma^{y}$ operations connecting the top and bottom edges of the system; this operator commutes with $H$ and returns opposite signs for the two ground states. 

Systems with physical edges are obviously attractive for the purposes of building a physical quantum computer, but the edges present an important challenge to the shadow lattice architecture detailed in this work. In the torus geometry, bit flips always create or annihilate anyons in pairs, and the combination of the attractive ranged interaction and secondary repair qubits make it rare for anyons to wander apart before they are annihilated. However, a $\sigma^y$ error occurring on the top or bottom edge is capable of creating a single anyon (at an energy cost $V$ instead of $2V$), which can wander freely if it manages to escape the edge before a primary repair qubit annihilates it, as there is no partner anyon remaining in the system to pull it back. To remedy this, we introduce an external potential $\delta V \of{\mathbf{r}}$ acting on the anyons by locally adjusting $V_0$ (or a compensating ``background charge" of further boson source terms) from plaquette to plaquette or star to star.

\begin{figure}
\includegraphics[width=3.25in]{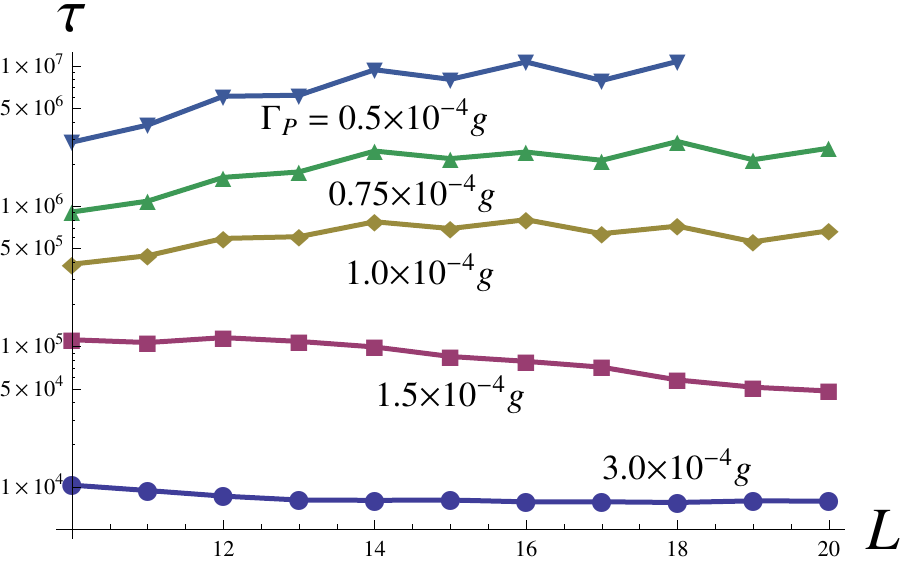}
\caption{Quantum state lifetime (in units of $g^{-1}$) of an $L \times L$ plaquette implementation of our system with physical edges, using the parameters described in the text and calculated using the defect tracking algorithm in Appendix A. A linear potential gradient $\delta V \of{\mathbf{r}} = - \omega_{9} \abs{y-L/2}$ is applied to recapture single anyons created at either edge;  $\omega_9$ is the lowest shadow lattice energy in Table I and $y$ is the vertical coordinate. The curves are computed from primary error rates $\Gamma_P = \cuof{ 3, 1.5, 1, 0.75, 0.5} \times 10^{-4} g$ (bottom to top curves) as in FIG.~\ref{datafig}, again using $J = 3g$, $\mu/J = - 0.1$ and the shadow lattice parameters in Table I. The even-odd effects seen at lower error rates stem from $y=L/2$ lying on or between plaquette centers. To mitigate further edge effects, the ranged interaction $U \of{\mathbf{r}}$ was calculated assuming that the $L \times L$ primary lattice was embedded at the center of a much larger ($100 \times 100$) resonator array. The parameters chosen here almost certainly do not represent the optimal choice for $\delta V \of{\mathbf{r}}$ or the shadow lattice energies. To reduce computation time, the values of $\tau$ with $\Gamma_P = 0.75 \times 10^{-4} g$ and $0.5 \times 10^{-4}$ were computed from 400 error simulations per data point (compared to 900 for the other values) and thus have a slightly higher uncertainty. }\label{edgedata}
\end{figure}

There is of course no unique choice for $\delta V \of{\mathbf{r}}$, and we did not attempt to find its optimal functional form in this work. Rather, in the simulations presented in FIG.~\ref{edgedata}, we chose the simple linear form $\delta V \of{\mathbf{r}} = - \omega_{9} \abs{y-L/2}$, where $\omega_9$ is the lowest shadow lattice energy in Table I, $y$ is the vertical coordinate, $L$ is the size of the system, and the distance between neighboring plaquette centers is 1. For even $L$, $y=L/2$ lay between two plaquette centers, whereas for odd $L$, $y=L/2$ lay at the center of a row of plaquettes, leading to the visible even-odd effects in FIG.~\ref{edgedata}. A third set of primary repair qubits with parameters $\Omega = 0.25 g$, $\Gamma_S = 0.3 g$ and $\omega = 30 g - L \omega_9/2$ was added at the edges to eliminate single anyons, and the energies of each primary refilling qubit were locally adjusted to take the changing values of $V$ into account. An additional potential gradient for the star terms would be applied along the $x$ axis to handle the other error channel. The above choice for $\delta V \of{\mathbf{r}}$ allows the shadow lattice to pull anyons toward either edge and thus recapture single anyons, while being weak enough that nearby anyons (such as pairs created near $y=L/2$) can still be efficiently drawn back together. The external potential gradient naturally reduces the ``useful" range of the interaction potential, and while the final lifetimes we obtain are generally around a factor of ten smaller than in the periodic geometry, the scaling of $\tau$ versus $\Gamma_P$ clearly demonstrates multistage passive error correction. Finely tuning the shadow lattice and shape of the potential could likely produce dramatic improvements in $\tau$ relative to the values presented here; we merely sought to demonstrate that shadow lattice error correction can handle edges in addition to the periodic geometry considered earlier in the work. 

\section{Discussion}\label{dissec}

We have presented a relatively simple mechanism by which long chains of quantum logical errors (of any type) can be passively self corrected in a topological qubit array. Our system is explicitly evaluated in a rotating frame and the resulting infinite temperature bath coupling is balanced by a much stronger coupling to a colored, zero temperature bath (the shadow lattice), leading to many stages of effective quantum error correction. We have thus demonstrated that long-ranged self correction of quantum errors could be implemented in device arrays, and while the question of whether or not two- or three-dimensional self correcting quantum code is possible in the thermodynamic limit and in a realistic device model remains open (and we suspect the answer to be negative), in finite size systems quantum self correction is a real experimental possibility. We must however point out that we have not considered the effect of \textit{static} error sources, such as random variations in the individual qubit parameters, and the constraints on the magnitudes of these fluctuations could be quite stringent if passive error correction is to be realized.

While Kitaev's toric code is the simplest commuting stabilizer model, our passive error correction mechanism could be straightforwardly generalized to other similar Hamiltonians, such as topological color codes \cite{bombin2010} and string-net models \cite{levinwen2005} or related quantum loop gases. While generally requiring higher order interactions (and thus being more difficult to realize), some of these systems have non-abelian anyon excitations \cite{nayaksimon} and could be more efficient from a computational point of view than the toric code. In principle, ranged interactions and a shadow lattice could also be used to induce passive recombination of quasiholes in both abelian and non-abelian bosonic fractional quantum Hall states \cite{kapithafezi2014}. However, when ranged interactions are added to the Kapit-Mueller Hamiltonian \cite{kapitmueller,liubergholtz2014} the ground state wavefunctions are no longer known exactly, and the fact that the component particles are mobile forbids the use of the similarity transformation we used to decouple the toric code from the propagating bosons. Proving quantum self correction in such systems would thus be a much more difficult undertaking than in commuting stabilizer models such as the one studied in this work.

We finally wish to point out that appropriately implemented, the shadow lattice construction is compatible with algorithmic measurement-based error correction schemes such as the surface code. When combined with our passive error correction mechanism, even a rudimentary surface code implementation with a long cycle time could still produce logical state lifetimes vastly longer than those reported here, and could overcome the saturation effects we observed when the system became too large compared to the interaction range. Likewise, a comparatively primitive shadow lattice implementation, using only primary error correction qubits and no propagating bosons or ranged interactions, could still produce dramatic improvements when used to supplement a measurement-based approach. Hybrid quantum error correction, which combined both active and passive elements, could prove to be an invaluable tool for constructing a future quantum computer, and would be an extremely fruitful topic for future research. 

\section{Acknowledgments}

We would like to thank Jens Koch, Mohammad Hafezi, Neil Robinson, David Schuster and Curt von Keyserlingk for useful discussions. This material is based on work supported by EPSRC Grant Nos. EP/I032487/1 and EP/I031014/1, and Worcester College of the University of Oxford.

\section{Appendix A: Defect tracking algorithm}

The algorithm we used to compute state lifetimes is fairly simple and computationally efficient, though an exponential increase in the state lifetime necessarily requires an exponential increase in the computational cost of the simulation. In all of our simulations, we only considered one type of quantum error (plaquette violations) as the physics of the two error types is identical and they do not interfere with each other. 

For systems with periodic boundary conditions, our calculation worked as follows. We initialize the system (with $N$ spins and $N/2$ plaquettes) in a ground state, so that all $Q_{\diamondsuit j}$ operators evaluate to zero. We also begin with an integer $k=1$ and an empty list called the edge list. For each timestep $dt$ (chosen to be small enough that finite size corrections are negligible), we compute the error probability $dt \of{\Gamma_P + \Gamma_{Ri}}$ for each spin $i$ in the primary lattice, where $\Gamma_P$ is the global, energy-independent error rate and $\Gamma_{Ri}$ is the correction from the shadow lattice, which depends on the values of all the plaquette operators $Q_{\diamondsuit j}$. We then generate $N$ random numbers on the interval between 0 and 1, and when one of these numbers is less than the corresponding error probability, we flip that spin. If the two plaquettes adjoining that spin are in the ground state ($Q_{\diamondsuit j}=0$), we flip them to $Q_{\diamondsuit j}=1$, label them both by $k$ and increment $k$ by one so that the label is not reused later. If the spin sits on the top edge of the system (the periodic boundary-- \textit{not} a physical edge) we append $k$ to the edge list. Similarly, if one of the adjoining plaquettes is empty and the other contains an anyon, we exchange the values of the two plaquettes, keeping track of the anyon's label $p$, and appending $p$ to the edge list if the flipped spin lies on the edge. 

If both plaquettes adjoining the flipped spin contain anyons, we annihilate them and check for an error. If the anyons have different labels $p$ and $q$, we remove them and then change all other instances of the label $q$ (in the system and in the edge list) to $p$. If the anyons share the same label $p$, no relabeling is needed. If the flip occurs at the edge, we append the label $p$ to the edge list. If the two anyons shared the same label before annihilation, we then look at the edge list, and if $p$ occurs an \textit{odd} number of times, then the anyons have traversed a non-contractible loop and a quantum error has occurred. We denote the current time $t$ and abort the calculation. If an error is not detected in that timestep, we simply repeat the process, incrementing $t+dt$ until an error is detected. By repeating the simulation many times, we can compute an average lifetime $\tau$, the lifetime of the quantum logical state encoded in the system's degeneracy. The relabeling upon annihilation is used to track anyon fusion, as a chain of fusions is topologically equivalent to moving a single anyon. It also ensures that the same label $p$ will be present at exactly two plaquettes in the lattice, until it is removed through annihilation. As the anyons are identical and indistinguishable defects, it is only by motion and chains of fusions that a noncontractible loop can form, leading to a quantum logic error. Note that there is no way to deform a loop of spin flips so that an even number of edge crossings becomes an odd number, so the parity of the number of edge crossings is the correct topological quantity to track.

For systems with physical edges, we kept two edge lists, the top and bottom lists. In this case, anyons can be created singly through a spin flip at the top or bottom edge; in each case a new label $k$ is generated. When a pair of anyons fuses together, or when an anyon fuses with an edge, we perform an appropriate relabeling if needed (taking care to replace any instances of the replaced label in each edge list), and then check both edge lists-- if the label $p$ occurs in \textit{both} edge lists, then an error has occurred. An error in this case corresponds to a string which joins the two edges, and as there is no way to deform the string so that it couples to only one edge without creating or annihilating anyons, it is a topologically invariant operation. 

We wish to point out that this error tracking procedure is different from most traditional quantum error correction simulations in commuting stabilizer models. In those cases (where there is no physical Hamiltonian, and the plaquette/star constraints are implemented through repeated measurements), at each timestep a map of all plaquette and star violations is constructed, and an algorithm is used to generate a set of additional operations which return the system to its ground state. The code fabric is thus cleared of anyons at every timestep, with a logical error occurring whenever the process of removing anyons enacts a topologically nontrivial chain of operations. In our system, anyons are not cleared at each timestep, and will instead persist until the shadow lattice or further random errors eliminate them. Our simulation only detected a logical errors upon the annihilation of anyons into a noncontractible loop, and we would have computed different lifetimes if we had instead detected logical errors by finding the time at which an algorithmic correction would lead to a logical error upon application (note of course that this algorithmic correction would not be enacted in the simulation). However, this condition would only occur when anyons had already become well-separated, from which point the average time to error scales as $\Gamma_{P}^{-1}$ and is not suppressed by the shadow lattice. As our computed lifetimes were much longer than this, we expect that choosing this alternative criteria for logical error correction would produce at most an $O \of{1}$ correction to our values for $\tau$.

\section{Appendix B: Three body gadget}

\begin{figure*}
\includegraphics[width=6in]{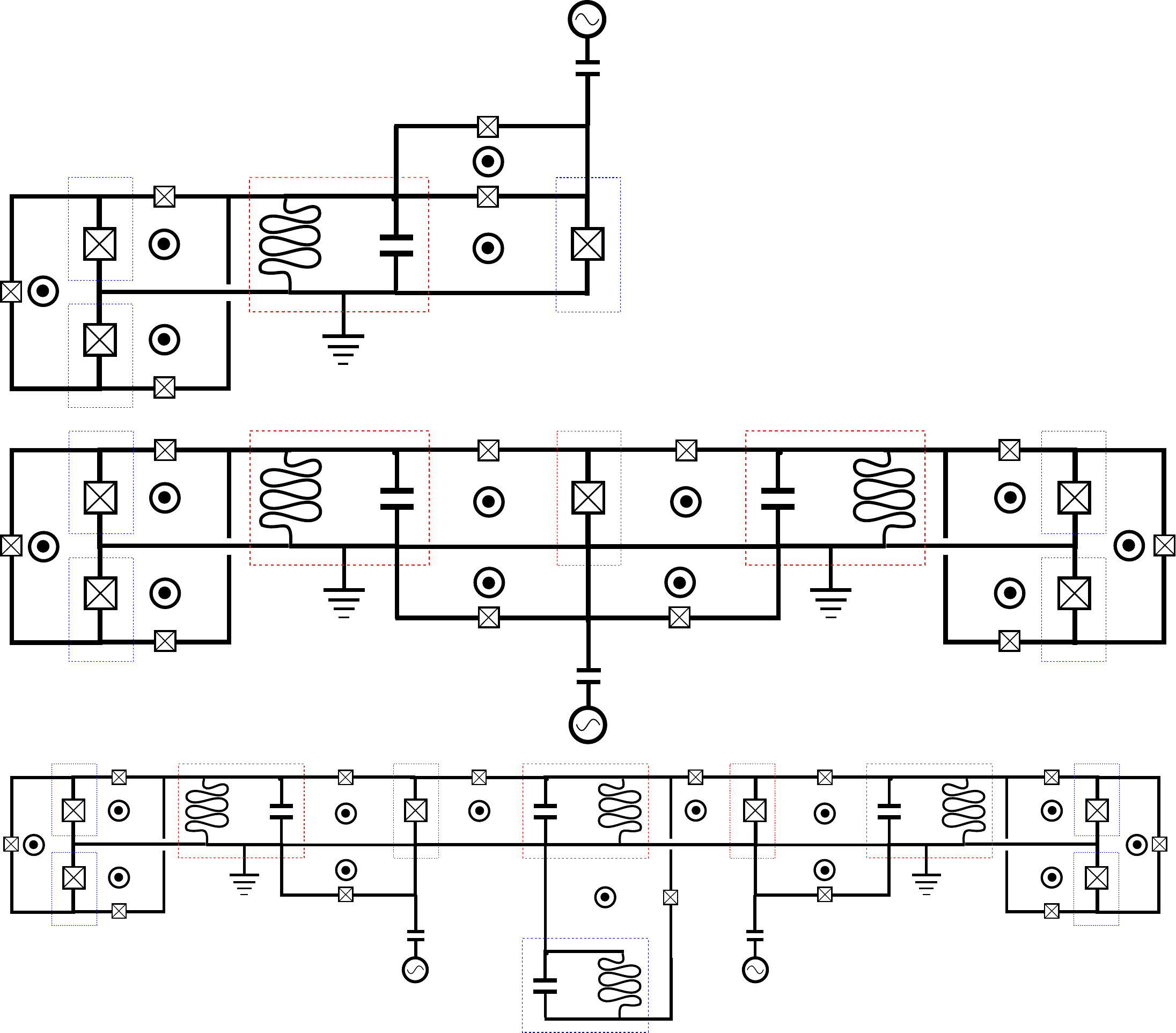}
\caption{(Color online) 3-, 4- and 5-body gadgets, implemented in transmon qubits. The Josephson junctions in blue boxes are the qubits of the primary lattice, and the circuits in red boxes are high-energy gadget degrees of freedom which are integrated out to produce 3- and 4-body couplings in perturbation theory. Appropriately tuned, the 5-body gadget will generate both 4- and 5-body terms ($-\sigma_{1}^{x} \sigma_{2}^{x} \sigma_{3}^{x} \sigma_{4}^{x} \of{ J_4' + J_5 \of{a+a^\dagger}}$), implementing the qubit Hamiltonian Eq.~\ref{fullH}. All devices share a common (bridged) ground and are coupled by small Josephson junctions with time-dependent flux biases, as described in Eq.~\ref{biases} in Appendix B. }\label{3gadgetfig}
\end{figure*}

To engineer a physical Hamiltonian for commuting stabilizer models, we need to devise a circuit that generates a continuous 3-body gate, with all unwanted 2-body interactions eliminated perturbatively through counterterms. Circuits of this type are commonly referred to as perturbative gadgets \cite{kempekitaev,jordanfarhi2008,ockoyoshida}, and a generic 3-body gadget--- a term which takes the form $\sigma_{i}^{a} \sigma_{j}^{b} \sigma_{j}^{c}$ after all high energy degrees of freedom are integrated out in perturbation theory--- is sufficient, in principle, for engineering arbitrary $n$-body couplings, since they can be easily chained together to generate higher order couplings. There are many possible gadget constructions, and precise details of the gadgets are not important so long as the low energy Hamiltonian reduces to Eq.~\ref{fullH}, though in a real experiment additional circuitry might be needed to correct errors that arise from unwanted transitions in the internal (gadget) degrees of freedom. We will now propose one such gadget, which uses superconducting qubits parametrically coupled \cite{Bergeal:2010iu,Anonymous:2013bq} to a resonator to generate 3-body interactions. 

\begin{figure}
\includegraphics[width=3.25in]{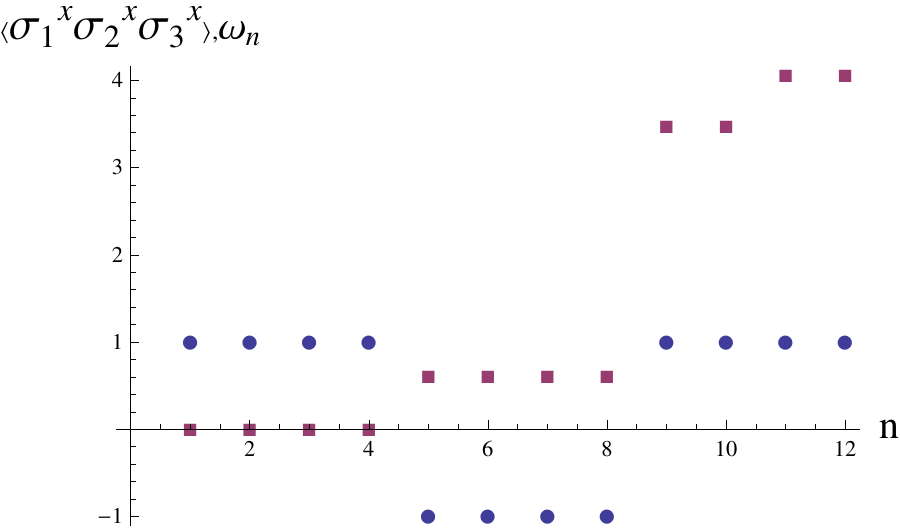}
\caption{(Color online) Eigensystem of the 3-body gadget derived from the construction in FIG.~\ref{3gadgetfig}. The blue circles plot the eigenvalues $\sigma_{1}^{x} \sigma_{2}^{x} \sigma_{3}^{x}$ of the eigenstates and the purple squares plot the energy in units of $\kappa$. The degeneracy can be made exact by careful tuning; values used in this simulation were (all in units of $\kappa$) $\kappa_3 = 1, \omega_R - \nu = 4, \kappa_c \simeq 0.64$ and $c_3 \simeq 0.3$. There are eight low-lying states, with the excited states all separated from them by a large excitation energy and thus inaccessible in the low-energy limit.}\label{3gadget-vals}
\end{figure}

We consider a circuit of 3 superconducting qubits, coupled to a central resonator through flux-biased Josephson junctions, as shown in FIG.~\ref{3gadgetfig}. All three qubits are charge-insensitive devices (such as transmon, flux or fluxonium qubits) and have excitation energies $\omega_{Qj}$; the resonator has excitation energy $\omega_R$. The Josephson energy $E_J'$ of these couplings is small compared to the internal Josephson energy $E_J$ of the qubits and compared to $\omega_R$. We will assume that the nonlinearities of all three qubits are  large compared to all other couplings, so that we can treat them as simple spin-$\frac{1}{2}$ degrees of freedom. Qubits 1 and 2 are coupled to each other through another small Josephson junction with energy $E_J''$. We let the rest frame basis of the qubits be $\sigma^z$. An oscillating voltage is applied to qubit 3 through a capacitive coupling, and qubit 3 is coupled to the resonator through two Josephson junctions with different flux biases. For charge noise-free superconducting qubits operated at a flux symmetry point, we make the operator identifications $Q \to q \sigma^{y}$, $\sin \phi \to s \sigma^x$ and $\cos \phi \to c + d \sigma^z$. For resonators, we have $\sin \phi \to s' \of{a + a^\dagger}$ and $\cos \phi \to c' + d' a^\dagger a + f \of{a^\dagger a^\dagger + a a }$. Of course, all terms which change the rest frame occupation of qubits or the resonator become rapidly oscillating in the rotating frame, and will require drive fields to cause transitions.

We choose the following flux biases for the qubit-resonator and qubit-qubit couplings: 
\begin{eqnarray}\label{biases}
\frac{\Phi_{1R} \of{t}}{\Phi_0} &=& \sqof{\frac{\pi}{2} + p_{1R} \of{\cos \of{\omega_{Q1} + \nu} t + \cos \of{\omega_{Q1} - \nu } t} } \nonumber \\
\frac{\Phi_{2R} \of{t}}{\Phi_0} &=& \sqof{\frac{\pi}{2} + p_{2R} \of{\cos \of{\omega_{Q2} + \nu} t + \cos \of{\omega_{Q2} - \nu } t} } \nonumber \\
\frac{\Phi_{3Ra} \of{t}}{\Phi_0} &=& \sqof{\frac{\pi}{2} + p_{3R} \cos \of{\frac{\omega_{Q3} }{2} +  \nu} t  }  \\
\frac{\Phi_{3Rb} \of{t}}{\Phi_0} &=& \sqof{\frac{\pi}{2} + p_{3R} \cos \of{\frac{\omega_{Q3} }{2} -  \nu} t  } \nonumber \\
\frac{\Phi_{12} \of{t}}{\Phi_0} &=& \sqof{\frac{\pi}{2} + p_{12} \of{\cos \omega_{12+} t + \cos \omega_{12-} t} }. \nonumber
\end{eqnarray}
Here, $\nu$ is close but not equal to $\omega_R$, $p_{1R} = p_{2R}$, $\omega_{12\pm} = \omega_{Q1} \pm \omega_{Q2}$, all $p$ coefficients are small compared to 1, and $\Phi_0 = h/2e$ is the superconducting flux quantum. The oscillating voltage applied to qubit 3 has the form $V \sin \omega_{Q3} t$. We move to a frame where the qubits each rotate at $\omega_{Qj}$ and the resonator rotates at $\nu$, so that $\sigma_{j}^{\pm} \to \sigma_{j}^{\pm} e^{\mp i \omega_{Qj} t }$ and $a^\dagger \to a^\dagger e^{-i \nu t}$. Expanding the junction couplings $- E_{J}' \cos \of{\phi_j - \phi_R - \Phi_{jR} \of{t}/\Phi_0 }$ to lowest nontrivial order in $p$ and keeping only the terms which are time independent in the rotating frame, we obtain
\begin{eqnarray}
H_{3} &=& \of{\omega_R -\nu} a^\dagger a - \kappa \of{ a + a^\dagger} \of{\sigma_{1}^{x} + \sigma_{2}^{x} } + \kappa_{c} \sigma_{1}^{x} \sigma_{2}^{x} \nonumber \\
& & + c_{3} \sigma_{3}^{x} - \kappa_{3} \sigma_{3}^{x} \of{a a + a^\dagger a^\dagger}.
\end{eqnarray}
Appropriate phase shifts of the flux signals lead to $\sigma^y$ terms instead of $\sigma^x$ terms. We now want to integrate out the resonator by treating $H_3$ perturbatively in $\kappa/ \of{\omega_R - \nu}$. We carefully choose $\kappa_c$ and $c_3$ to eliminate the second order term $\sim 2 \frac{\kappa^{2}}{\omega_R - \nu} \sigma_{1}^{x} \sigma_{2}^{x}$ and the third order term $\sim 8 \frac{\kappa^{2} \kappa_{3}}{\of{\omega_R - \nu}^{2} } \sigma_{3}^{x}$. The precise coefficients of $\kappa_c$ and $c_3$ that lead to exact cancellations in the low-energy Hamiltonian can easily be determined numerically. After integrating out the resonator, we arrive at the final Hamiltonian $H = -J_3 \sigma_{1}^{x} \sigma_{2}^{x} \sigma_{3}^{x}$, where $J_3 \simeq \frac{4 \kappa^{2} \kappa_{3}}{\of{\omega_R - \nu}^{2} }$ and with all lower-order terms eliminated by the counterterms $\kappa_c$ and $c_3$. We demonstrate this for an example construction in FIG.~\ref{3gadget-vals}; the degeneracies of the four ground states with $\sigma_{1}^{x} \sigma_{2}^{x} \sigma_{3}^{x} = 1$ and of the four excited states with $\sigma_{1}^{x} \sigma_{2}^{x} \sigma_{3}^{x} = -1$ can both be made exact through careful tuning of $\kappa_c$ and $c_3$. The coefficient $J_3$ can actually be made fairly large while still remaining in the perturbative regime; for realistic superconducting devices, $\kappa/\hbar = 2 \pi \times 100$MHz is a readily achievable 2-body coupling, and for $\omega_R - \nu = 4 \kappa$ and $\kappa= \kappa_3$, this leads to a continuous three-body coupling of strength $J_3 /\hbar \simeq 2\pi \times 30$MHz, as shown in the figure. The ultimate limitations on the coupling strengths are the qubit nonlinearities, which must all be large compared to the $\kappa_j$ so that the qubits can be treated as simple spins.

\begin{figure}
\includegraphics[width=3.25in]{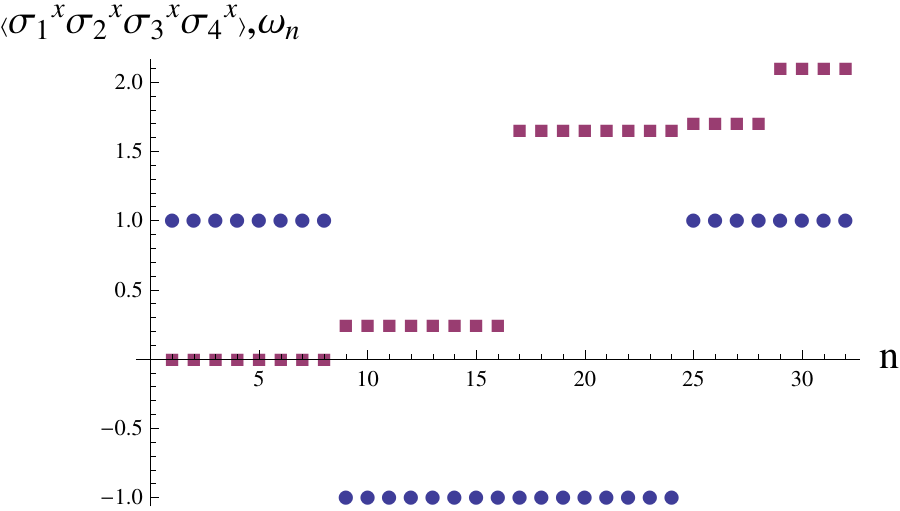}
\caption{(Color online) Eigensystem of the 4-body gadget constructed from chaining together 3-body gadgets. The low-energy Hamiltonian is $H = - J_{4} \sigma_{1}^{x} \sigma_{2}^{x} \sigma_{3}^{x} \sigma_{4}^{x} $, where $J_4 \simeq 2 J_{3}^{2}/\of{\omega_{Q5}-\nu'}$. The blue circles plot the eigenvalues $\sigma_{1}^{x} \sigma_{2}^{x} \sigma_{3}^{x} \sigma_{4}^{x}$ of the eigenstates and the purple squares plot the energy in units of $\kappa$. Higher order interactions can be also generated through this mechanism. Here the counterterm $\kappa_c$ is identical to the 3-body case, $c_5 = 2 c_3$, and $\omega_Q - \nu' = 2 \kappa$.}\label{4gadgetfig}
\end{figure}

We can chain these terms together to generate higher order interactions as well. Consider the second device shown in FIG.~\ref{3gadgetfig}, where two of the three-body gadgets share a central qubit, with excitation energy $\omega_{Q5}$ that is somewhat higher than the drive field $\nu'$ applied to it. The Hamiltonian for this device is
\begin{eqnarray}
H_4 &=& \of{\omega_R -\nu} \of{ a_{1}^\dagger a_{1} + a_{2}^\dagger a_{2} } - c_5 \sigma_{5}^{x} + \frac{\omega_{Q5} - \nu'  }{2} \sigma_{5}^{z} \\
& & - \kappa \sqof{ \of{ a_{1} + a_{1}^\dagger} \of{\sigma_{1}^{x} + \sigma_{2}^{x} } + \of{ a_{2} + a_{2}^\dagger} \of{\sigma_{3}^{x} + \sigma_{4}^{x} } } \nonumber \\
& & + \kappa_{c} \of{  \sigma_{1}^{x} \sigma_{2}^{x} + \sigma_{3}^{x} \sigma_{4}^{x} } \nonumber \\
& &  - \kappa_{5} \sigma_{5}^{x} \of{a_{1} a_{1} + a_{1}^\dagger a_{1}^\dagger + a_{2} a_{2} + a_{2}^\dagger a_{2}^\dagger}. \nonumber 
\end{eqnarray}
We now integrate out qubit 5 and the two resonators. With properly chosen counterterms, we arrive at $H = - J_{4} \sigma_{1}^{x} \sigma_{2}^{x} \sigma_{3}^{x} \sigma_{4}^{x} $, where $J_4 \simeq 2 J_{3}^{2}/\of{\omega_{Q5}-\nu'}$, with $J_3$ given by the coefficient of the 3-body gadget. We demonstrate the eigensystem of an example of this device in FIG.~\ref{4gadgetfig}. 

A 5-body gadget could be constructed by chaining together three of the 3-body gadgets. Though this is not the simplest or most efficient way to generate a 5-body coupling, we begin with a 3-body gadget, with qubit 3 replaced by a resonator ($\sigma_{3}^{x} \to a + a^\dagger$). Additional terms could be added through the driven flux bias to eliminate couplings that are quadratic or higher order in the resonator's creation and annihilation operators $a$ and $a^\dagger$. We then give qubits 1 and 2 additional $\sigma^z$ terms (as done with qubit 5 in the 4-body gadget of the previous paragraph), and use each of them as the third qubit of an additional 3-body gadget, giving a total of six qubits and four resonators for the entire device. The resulting many-body coupling has the form $-\sigma_{1}^{x} \sigma_{2}^{x} \sigma_{3}^{x} \sigma_{4}^{x} \of{ J_4' + J_5 \of{a+a^\dagger}}$, engineering the basic qubit Hamiltonian in Eq.~\ref{fullH}.

\bibliography{SLbib,biblio,EC_bib}

\end{document}